\documentclass[journal]{IEEEtran}
\usepackage{cite}
\usepackage{amsmath,amssymb,amsfonts}
\usepackage{algorithmic}
\usepackage{graphicx}
\usepackage{textcomp}
\usepackage{xcolor}
\usepackage{booktabs}
\usepackage{multirow}
\usepackage{rotating}
\usepackage{makecell}
\usepackage{amssymb}
\usepackage{array}
\usepackage[inline]{enumitem}
\usepackage{adjustbox}
\usepackage{amsmath}
\usepackage{soul}
\usepackage[most]{tcolorbox}
\usepackage{xcolor}
\usepackage{lipsum}
\usepackage{balance}
\usepackage[table]{xcolor} 
\usepackage{colortbl}

\usepackage[hyphens]{url} 
\usepackage[breaklinks=true]{hyperref}

\newtcolorbox{rqanswerbox}{
  enhanced,
  colback=gray!15,
  colframe=gray!15,                       
  boxrule=0pt,                            
  borderline west={2pt}{0pt}{gray!70},   
  sharp corners,
  boxsep=4pt,
  left=4pt, right=4pt, top=4pt, bottom=4pt,
  before skip=1ex,
  after skip=1ex
}

\usepackage{ifthen}

\newboolean{showchangesformajor}
\setboolean{showchangesformajor}{false} 

\usepackage{xcolor}
\newenvironment{revision}{%
    \ifthenelse{\boolean{showchangesformajor}}%
        {\color{blue}}
        {}
}{%
    \ignorespacesafterend
}

\newcommand{\rev}[1]{\ifthenelse{\boolean{showchangesformajor}}{\textcolor{blue}{#1}}{#1}}

\newboolean{showchangesforminor}
\setboolean{showchangesforminor}{false} 

\usepackage{xcolor}
\newenvironment{revisionForMinor}{%
    \ifthenelse{\boolean{showchangesforminor}}%
        {\color{blue}}
        {}
}{%
    \ignorespacesafterend
}

\newcommand{\revForMinor}[1]{\ifthenelse{\boolean{showchangesforminor}}{\textcolor{blue}{#1}}{#1}}

\begin{document}

\title{Does AI Code Review Lead to Code Changes? A~Case~Study~of~GitHub~Actions}

\author{Kexin Sun, Hongyu Kuang, Sebastian Baltes, 
        Xin Zhou, He Zhang, Xiaoxing Ma, 
        Guoping Rong, Dong Shao, and Christoph Treude
\thanks{K. Sun, H. Kuang, X. Zhou, H. Zhang, X. Ma, G. Rong, and D. Shao are with State Key Lab for Novel Software Technology, Nanjing University, Nanjing, China (e-mail: kexinsun@smail.nju.edu.cn; khy@nju.edu.cn; zhouxin@nju.edu.cn; hezhang@nju.edu.cn; xxm@nju.edu.cn; ronggp@nju.edu.cn; dongshao@nju.edu.cn).}%
\thanks{S. Baltes is with University of Heidelberg, Heidelberg, Germany (e-mail: sebastian.baltes@uni-heidelberg.de).}%
\thanks{C. Treude is with School of Computing and Information Systems, Singapore Management University, Singapore (e-mail: ctreude@smu.edu.sg).}
\thanks{Corresponding author: Hongyu Kuang (e-mail: khy@nju.edu.cn).}}

\markboth{}{}


\maketitle

\begin{abstract}
AI-based code-review tools automatically review and comment on pull requests to improve code quality.
Despite their growing presence, little is known about their actual impact.
We present a large-scale empirical study of 16 popular AI-based code-review actions for GitHub workflows, analyzing more than 22,000 review comments in 178 repositories.
We investigate (1) how these tools are adopted and configured, (2) whether their comments lead to code changes, and (3) which factors influence their effectiveness.
We develop a two-stage LLM-assisted framework to determine whether review comments are addressed.
We then use interpretable machine learning to identify the influencing factors.
We found that while adoption is growing, its effectiveness varies widely.
Comments that are concise, contain code snippets, and are manually triggered, particularly those from hunk-level review tools, are more likely to result in code changes. 
These results highlight the importance of tool design and suggest directions for improving AI-based code review systems.
\end{abstract}


\section{Introduction}

Code review is a widely adopted practice in modern software engineering, playing a central role in identifying bugs, improving code quality, and facilitating knowledge transfer within teams~\cite{sadowski2018modern}. Traditionally, code reviews are performed manually through reviewing and commenting on code changes (submitted via pull requests) by developers~\cite{gousios2016work}. However, recent advances in generative AI (GenAI) have created opportunities to automate at least part of this process by generating human-like code review comments~\cite{lu2023llama}.

On platforms such as GitHub, developers can now integrate GenAI into their workflows via GitHub Actions~\cite{wessel2023github}. These AI-driven code review actions automatically analyse code changes and post suggestions as comments on pull requests.
Some of these actions are now available and popular on the GitHub Marketplace, promising to reduce developers' efforts and to increase review coverage~\cite{decan2022use}. Despite the growing presence, there is little empirical evidence of how these actions are actually used in practice and whether they can impact the development process meaningfully.

Previous research has investigated human code review practices in detail~\cite{bacchelli2013expectations}, including (1) factors that make review comments useful~\cite{bosu2015characteristics, turzo2024makes, rahman2017predicting}, (2) the granularity of feedback~\cite{lin2024leveraging}, and (3) developer responsiveness~\cite{macleod2017code}. In contrast, the dynamics of automated reviews powered by GenAI remain largely unexplored. Open questions include: Which AI-based tools are developers actually adopting? Do generated comments lead to meaningful code changes? Which characteristics influence whether such comments are taken seriously and acted upon?

This paper addresses these gaps through an empirical study of AI-based code review actions based on Large Language Models (LLMs) on GitHub. Our goal is to understand both the usage and the effectiveness of these tools, as well as the factors that influence their impact. Specifically, we propose the following three research questions:

\begin{description}[style=multiline, leftmargin=9mm]
\item[RQ1:] \emph{How are LLM-based code review actions adopted in GitHub repositories?}
\end{description}

This question is motivated by the proliferation of AI-based review tools on GitHub. Although many tools exist, their real-world adoption and configuration practices are not well understood. Studying adoption patterns allows us to assess the maturity and utility of these tools in practical software development workflows.

\begin{revision}
\begin{description}[style=multiline, leftmargin=9mm]
\item[RQ2:] \emph{To what extent do AI-generated review comments lead to code changes compared to human review comments?}
\end{description}

Our second question aims to understand whether AI-generated review comments can truly impact the code (i.e., lead to code changes) and how their effectiveness compares to human review comments. To enable large-scale analysis, we propose an LLM-based approach to automatically assess the validity of review comments and whether they had been addressed.
\end{revision}

\begin{description}[style=multiline, leftmargin=9mm]
\item[RQ3:] \emph{Which factors impact the likelihood that code review comments lead to code changes?}
\end{description}

Finally, we explore the contextual and technical factors that make comments more or less likely to result in code changes. By analyzing aspects such as comment content, contributor experience, and project characteristics, we aim at identifying patterns that can inform the design of more effective review~tools.

\begin{figure*}[t] 
	\centering
    \setlength{\abovecaptionskip}{3pt}
	\includegraphics[width=\textwidth]{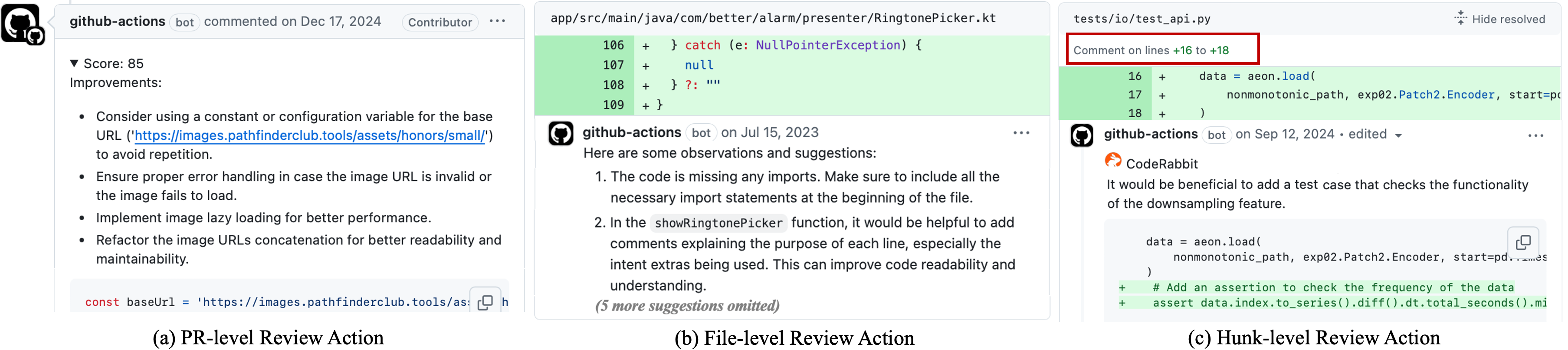} 
    \caption{Example comments from a \textbf{PR-level} review action (Integral-Healthcare/robin-ai-reviewer), a \textbf{file-level} review action (anc95/ChatGPT-CodeReview), and a \textbf{hunk-Level} review action (coderabbitai/ai-pr-reviewer).}
	\label{fig:action_example}
\end{figure*}

To answer these questions, we conducted a large-scale empirical study of 16 popular GenAI-based code-review actions on GitHub. We analyze their adoption in 178 repositories, collect more than 22,000 review comments, and develop a two-stage LLM-based classification framework to assess whether these comments are addressed. Finally, we use interpretable machine learning techniques to model and find out the factors that influence the effectiveness of comments.

Our contributions are as follows.
\begin{itemize}
    \item We provide the \textbf{first systematic study} of the \textbf{adoption and usage} of \textbf{AI-based code review} actions on GitHub, categorizing tools by their review granularity (i.e., PR, file, or hunk) and behaviour (e.g., action triggers). 
    \item We introduce an \textbf{LLM-assisted framework} for assessing whether code review comments are actionable and whether they have been \textbf{addressed}, achieving high accuracy compared to human annotations.
    \item We identify and explain the \textbf{factors} that influence whether \textbf{AI-generated comments} lead to \textbf{code changes}, offering design implications for code review tools. 
\end{itemize}

Our findings offer insights into the role of GenAI in software engineering that are relevant to researchers and practitioners interested in automated software quality assurance. 


\section{Background}

GitHub Actions is a built-in automation platform on GitHub that allows developers to define custom workflows using reusable components called ``\textit{actions}''.
This section uses \texttt{anc95/ChatGPT-CodeReview}, the most popular action in our dataset, as an example to demonstrate how AI-based code review can be integrated into a project’s workflow.

\begin{table*}[t]
  \centering
  \caption{Comparison of design choices for selected AI-based code review actions across each \textbf{granularity category} (ID-11: Integral-Healthcare/robin-ai-reviewer for \textbf{PR-level}, ID-1: anc95/ChatGPT-CodeReview for \textbf{file-level}, ID-3: coderabbitai/ai-pr-reviewer for \textbf{hunk-level}). LLM selection is the default model in the latest version.}
   \setlength{\tabcolsep}{2pt}
   \renewcommand{\arraystretch}{1.1}
   \resizebox{\textwidth}{!}{
    \begin{tabular}{c|
    >{\centering\arraybackslash}m{14em}|
                    >{\raggedright\arraybackslash}m{24em}|
                    >{\centering\arraybackslash}m{5em}|
                    >{\centering\arraybackslash}m{5em}|
                    >{\centering\arraybackslash}m{11em}|
                    >{\centering\arraybackslash}m{9em}}
    \hline
    \textbf{ID} 
    & \multicolumn{1}{c|}{\textbf{Action Description}} 
    & \multicolumn{1}{c|}{\textbf{Input Change Scope}} 
    & \textbf{Processing\newline{}Granularity} 
    & \textbf{LLM\newline{}Selection} 
    & \multicolumn{1}{c|}{\textbf{Prompt Context}} 
    & \multicolumn{1}{c}{\textbf{Commenting Format}} \\
     \hline
    11    
    & \textbf{PR-level code review} 
    & Changes between PR's base and latest commit. 
    & PR-level
    & gpt-4-turbo 
    & \textit{(Only diff, no additional)}
    & General comment \\
    \hline
    1     
    & \textbf{File-level code review} 
    & Optional based on PR event:\newline{}
    If \textit{synchronize}: 
    Only Changes in the latest commit.\newline{}
    If \textit{others}: 
    Changes between PR's base and latest commit. 
    & File-level 
    & gpt-4o-mini 
    & \textit{(Only diff, no additional)}
    & Inline comment \\
    \hline
    3     
    & \textbf{Hunk-level code review}\newline{}(also: generating PR descriptions, code change summaries, and direct interaction with the LLM) 
    & Changes between PR's the last reviewed commit and the latest commit. 
    & File-level 
    & gpt-4 
    & PR title; PR description; LLM's change summary 
    & Inline comment \\
     \hline
    \end{tabular}}%
  \label{tab:technical_choices}%
\end{table*}%

\begin{figure}[t] 
	\centering
    \setlength{\abovecaptionskip}{3pt}
	\includegraphics[width=0.46\textwidth]{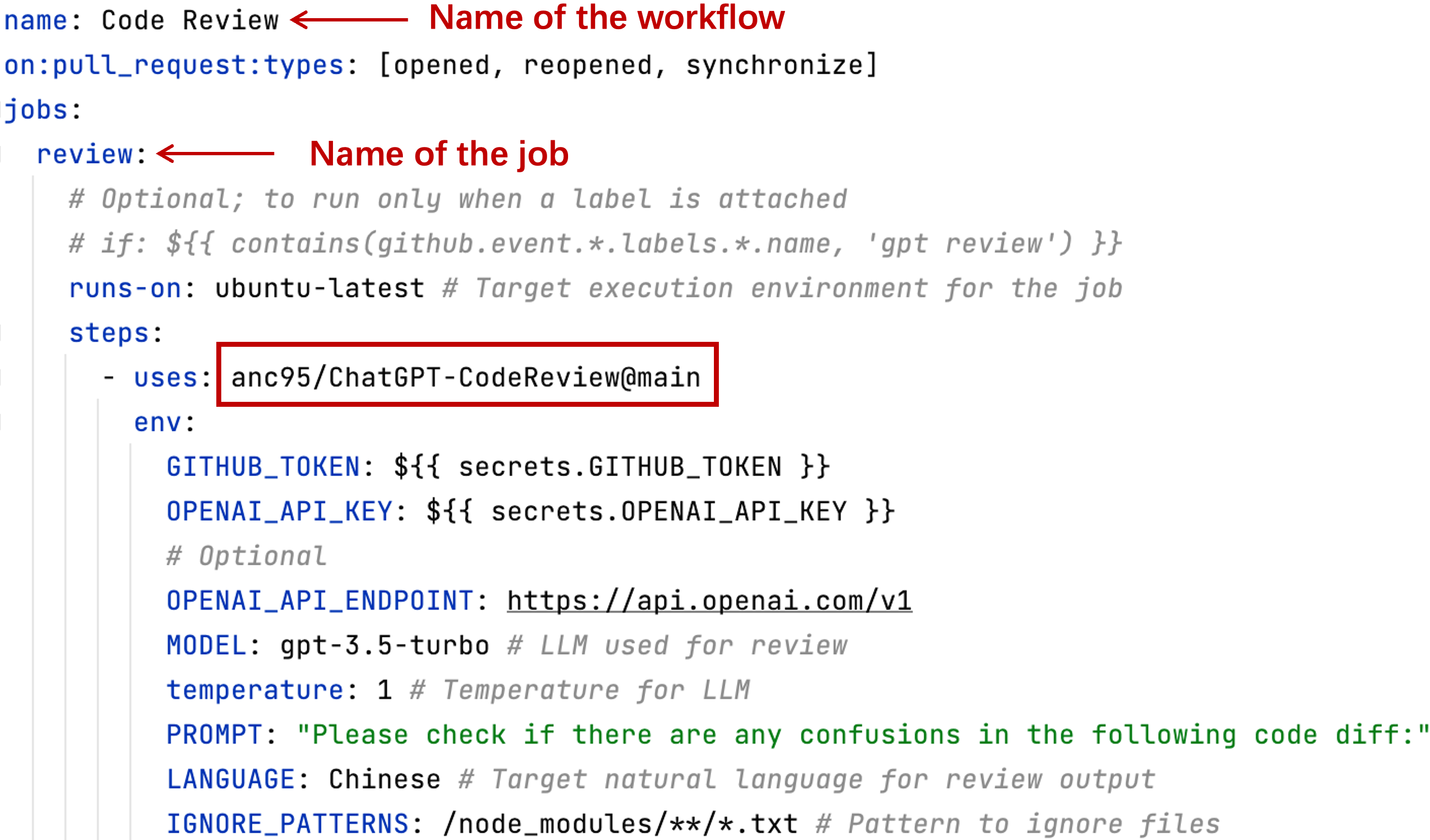} 
	\caption{An example of \textbf{configuring} an AI-based code review action.}
	\label{fig:configuring_example}
\end{figure}

\vspace{1ex}
\noindent
\textbf{How to Configure an Action:}
Developers can define and manage project workflows through YAML files in the project's \texttt{.github/workflows} directory.
As Fig.~\ref{fig:configuring_example} shows, in our example, we define a workflow named \texttt{Code Review}, containing a single job called \texttt{review} with one step that utilizes the main branch of \texttt{anc95/ChatGPT-CodeReview} for code analysis (\texttt{uses: anc95/ChatGPT-CodeReview@main}).
Using the keyword `\texttt{on}', we set this workflow to trigger automatically when pull requests are opened, reopened, or synchronized.
The trigger conditions can be further restricted using `\texttt{if}' conditions.
For example, as shown in the commented-out section of Fig.~\ref{fig:action_example}, we can restrict the workflow to trigger only on pull requests labeled with `\texttt{gpt review}'.
In our subsequent analysis, we refer to this conditional triggering as manual triggering because it requires developers to explicitly enable it.
The example action requires two mandatory parameters: \texttt{GITHUB\_TOKEN} for repository access and \texttt{OPENAI\_API\_KEY} for LLM authentication.
Additionally, we can customize its behavior through optional parameters,
such as which LLM to use (\texttt{MODEL}),
which prompt to apply (\texttt{PROMPT}),
and which natural language to comment in (\texttt{LANGUAGE}).

\vspace{1ex}
\noindent
\textbf{How the Action Works:}
Once the trigger conditions are satisfied (e.g., a pull request is opened), \texttt{ChatGPT-CodeReview} conducts a code review. 
It compares the pull request's base commit and the latest commit (the head commit at that time) to obtain the cumulative code changes, then reviews these changes file by file.
During this process, certain files may be skipped based on predefined ignore patterns, and the large change blocks may be divided according to the configured token limits. 
After the LLM analyzes a file's changes via the designed prompt, \texttt{ChatGPT-CodeReview} posts an inline comment to the pull request via the GitHub Action bot. This comment is typically attached to the last line of the final changed block within the file.
Fig.~\ref{fig:action_example}(b) provides an example of a review comment generated by \texttt{anc95/ChatGPT-CodeReview}.

\section{Study Setup}

To investigate AI-based code review on GitHub, we collected the 16 most popular actions as our study sample and conducted a categorization based on their review behaviors.
We provide an online appendix~\cite{appendix}, including our dataset, annotations, and scripts for LLM-assisted analysis of whether review comments have been addressed (for RQ2), and for the interpretation of how factors impact the likelihood of comment addressing (for RQ3).
We follow the guidelines proposed by Wagner et al.~\cite{llmguidelines-paper, baltes2025evaluationguidelinesempiricalstudies} for empirical studies in software engineering involving LLMs.


\vspace{1ex}
\noindent
\textbf{Action Selection:}
On January 7, 2025, we began our selection process by manually examining the names and descriptions of the top-ranked code review actions on the GitHub Marketplace (sorted by popularity).
From a review of approximately 240 actions, we identified an initial set of 20 candidates related to AI-driven code review.
We then carefully checked their project documentation and excluded four actions that focused on tasks other than comment generation (e.g., PR description generation).
This process yielded a final set of 16 relevant actions.
In this paper, we refer to each action by its repository's full name and report its star count in Table~\ref{table:llm_review_actions}.

\vspace{1ex}
\noindent
\textbf{Action Categorization:}
Although designed for the same purpose, these actions exhibit distinct review behaviors.
We group them into three categories based on the granularity of their feedback: \textbf{PR-level}, \textbf{file-level}, and \textbf{hunk-level}, as shown in Table~\ref{table:llm_review_actions} (``\textit{hunk}'' refers to a contiguous block of differing lines when comparing two versions of a file~\cite{hunk}).
These levels reflect increasing ability to provide context-specific comments.
\textbf{Hunk-level} review actions most closely resemble typical human code reviews.
They can provide targeted comments on specific code lines within files. 
For example, in Fig.~\ref{fig:action_example}(c), \texttt{coderabbitai/ai-pr-reviewer} suggests adding an assertion to check data frequency after lines 16-18 in \texttt{tests/io/test\_api.py}.
\textbf{File-level} review actions are not explicitly associated with specific lines but can still associate comments with the certain file.
As exemplified by the previously discussed \texttt{anc95/ChatGPT-CodeReview}, it attaches its reviews of the entire file's changes to the end of the overall file diff.
Consequently, developers must manually map suggestions to specific code blocks within files to comprehend the recommended modifications.
\textbf{PR-level} review actions represent the coarsest granularity, lacking file-specific comment linking.
Fig.~\ref{fig:action_example}(a) illustrates this with a comment from \texttt{Integral-Healthcare/robin-ai-reviewer}, which discusses changes spanning multiple files within a single review message.

\begin{table}[t]
\centering
\caption{Overview of 16 selected AI-based \textbf{Code Review} Actions.}
 \resizebox{0.48\textwidth}{!}{
\begin{tabular}{@{}c|c|c|c|c|c}
\hline
\multirow{2}[3]{*}{\textbf{ID}} & \multirow{2}[3]{*}{\textbf{Action}} & \multirow{2}[3]{*}{\textbf{Stars}} & \multicolumn{3}{c}{\textbf{Review Granularity}} \\
\cline{4-6}          &       &       
& \textbf{\shortstack{\\[-0.4ex]PR-\\level}}
& \textbf{\shortstack{\\[-0.4ex]File-\\level}}
& \textbf{\shortstack{\\[-0.4ex]Hunk-\\level}} \\
\hline
1  & \href{https://github.com/anc95/ChatGPT-CodeReview}{anc95/ChatGPT-CodeReview}        & 4.1k &        & \checkmark &        \\
2  & \href{https://github.com/mattzcarey/code-review-gpt}{mattzcarey/code-review-gpt}      & 1.7k &        & \checkmark &        \\
3  & \href{https://github.com/coderabbitai/ai-pr-reviewer}{coderabbitai/ai-pr-reviewer}     & 1.6k &        &   & \checkmark      \\
4  & \href{https://github.com/aidar-freeed/ai-codereviewer}{aidar-freeed/ai-codereviewer}    & 704  &        &   & \checkmark      \\
5  & \href{https://github.com/kxxt/chatgpt-action}{kxxt/chatgpt-action}             & 558  & \checkmark      &   &        \\
6  & \href{https://github.com/cirolini/genai-code-review}{cirolini/genai-code-review}      & 339  & \checkmark      &   &        \\
7  & \href{https://github.com/truongnh1992/gemini-ai-code-reviewer}{truongnh1992/gemini-ai-code-reviewer} & 101 &   &   & \checkmark      \\
8  & \href{https://github.com/feiskyer/ChatGPT-Reviewer}{feiskyer/ChatGPT-Reviewer}       & 70   & \checkmark      &   &        \\
9  & \href{https://github.com/adshao/chatgpt-code-review-action}{adshao/chatgpt-code-review-action} & 69 & \checkmark      &   &        \\
10 & \href{https://github.com/tmokmss/bedrock-pr-reviewer}{tmokmss/bedrock-pr-reviewer}     & 61   &        &   & \checkmark      \\
11 & \href{https://github.com/Integral-Healthcare/robin-ai-reviewer}{Integral-Healthcare/robin-ai-reviewer} & 59 & \checkmark &   &        \\
12 & \href{https://github.com/presubmit/ai-reviewer}{presubmit/ai-reviewer}           & 53   &        &   & \checkmark      \\
13 & \href{https://github.com/gvasilei/AutoReviewer}{gvasilei/AutoReviewer}           & 42   &        & \checkmark &        \\
14 & \href{https://github.com/unsafecoerce/chatgpt-action}{unsafecoerce/chatgpt-action}     & 39   &        &   & \checkmark      \\
15 & \href{https://github.com/magnificode-ltd/chatgpt-code-reviewer}{magnificode-ltd/chatgpt-code-reviewer} & 27 &   & \checkmark &        \\
16 & \href{https://github.com/ca-dp/code-butler}{ca-dp/code-butler}               & 23   & \checkmark      &   &        \\
\hline
\end{tabular}}
\label{table:llm_review_actions}
\end{table}

\begin{table*}[t]
\centering
\caption{Categories of \textbf{Optional Configuration Parameters} in AI-based Code Review Actions with Descriptions and Examples}
\resizebox{\textwidth}{!}{
\begin{tabular}{c|l}
\hline
\textbf{Configuration Category} & \multicolumn{1}{c}{\textbf{Description and Example}} \\
\hline
\textbf{Task Triggers \& Modes} & Defines when and how the action executes. E.g., \texttt{review\_per\_file} in feiskyer/ChatGPT-Reviewer (enables file-level review if true). \\
\textbf{Input Settings} & Excludes/includes specific files or limits code diff size. E.g., \texttt{IGNORE\_PATTERNS} in aidar-freeed/ai-codereviewer (specifies file exclusion patterns). \\
\textbf{LLM Service Settings} & Defines parameters for connecting to the LLM service. E.g., \texttt{OPENAI\_API\_BASE} in Integral-Healthcare/robin-ai-reviewer (URL of the OpenAI API interface). \\
\textbf{LLM Selection} & Specifies the LLM used for review. E.g., \texttt{MODEL} in anc95/ChatGPT-CodeReview (name of the LLM). \\
\textbf{LLM Hyperparameters} & Configures generation behavior (e.g., randomness, length). E.g., \texttt{temperature} in feiskyer/ChatGPT-Reviewer (controls output randomness). \\
\textbf{Prompt Customization} & Defines the instructions given to the LLM. E.g., \texttt{PROMPT} in ca-dp/code-butler (custom instructions for review). \\
\textbf{Prompt Context Augmentation} & Additional context for prompt construction. E.g., \texttt{NLP\_LANGUAGE} in anc95/ChatGPT-CodeReview (specifies output language). \\
\textbf{Output Settings} & Controls review output formatting or logic. E.g., \texttt{review\_comment\_lgtm} in coderabbitai/ai-pr-reviewer (posts comments even if no issues are found if true). \\
\textbf{Others} & Miscellaneous settings unrelated to the core review process. E.g., \texttt{debug} in unsafecoerce/chatgpt-action (enables debug mode if true). \\
\hline
\end{tabular}}
\label{tab:config_categories}
\end{table*}

Beyond review granularity, actions also differ in how they implement the review process.
These differences include the input change scope (which commits are compared), the processing granularity (how the changes are segmented), the LLM selection (which model is used), the prompt context (what additional information is provided), and the commenting format (how the review is delivered).
To illustrate the variation, we present an example action for each of the three granularity categories in Table~\ref{tab:technical_choices}.
A full comparison of 16 actions can be found in our online appendix.
We observe that 6 out of the 16 actions incorporate related tasks to support code review, such as \texttt{coderabbitai/ai-pr-reviewer}'s PR summarization and interactive feedback.
For the input scope, 11 actions review the full base-to-latest commit diff on each run, while 3 actions (e.g., \texttt{coderabbitai/ai-pr-reviewer}) track reviewed commits to avoid redundancy, and 2 actions, including \texttt{anc95/ChatGPT-CodeReview}, dynamically adjust the input scope based on PR events or pre-configured modes.
Processing granularity does not always align with the review granularity.
For instance, \texttt{coderabbitai/ai-pr-reviewer} submits the file-level diff into the LLM in one request, but can leverage prompt constraints (e.g., ``\textit{Comment with exact line number}'') to extract multiple line-specific comments.
Most actions (13/16) rely on OpenAI's model family~\cite{OpenAI} for review, with the remaining using alternatives such as Gemini~\cite{Gemini} or Claude~\cite{Claude}.
In addition, 11 actions enrich the prompts with contextual information, such as the PR title, description, or programming language, alongside code changes.
Finally, actions offering file- or hunk-level review typically post feedback as inline comments (usually for detailed code review), while PR-level actions post as PR general comments (usually for broad discussion).
These varying design decisions will shape the developer experience and responses to the generated code review comments.

\section{Analyses \& Results}

Based on the 16 collected actions, in this section, we detail the methodology and results for each of our research questions.

\subsection{\textbf{RQ1:} How are LLM-based code review actions adopted in GitHub repositories?}

\vspace{1ex}
\noindent
\textbf{Approach:}
To answer this question, we examined the adoption of these actions by measuring their usage across repositories, pull requests, and comments. Furthermore, we analyzed their configuration settings to understand how developers adapt them to their specific project needs.

We utilized the GitHub REST API to collect relevant data.
First, we identified repositories containing configuration files that invoke the target actions.
Specifically, we retrieved the \texttt{.yml} and \texttt{.yaml} files in \texttt{.github/workflows} directories of GitHub repositories that explicitly referenced the target actions (i.e., containing ``\texttt{uses:ACTION}'' in uncommented sections).
These files were then grouped by repository. To ensure a unique mapping between repository, configuration file, and action, we filtered out 19 repositories that referenced multiple target actions or the same action across different files.
Following prior work~\cite{DBLP:conf/icse/ReichM23}, our analysis was limited to repositories with at least 50 PRs to ensure a minimum level of project maturity and popularity of PRs.
Next, we quantified action-generated comments by examining PRs in these repositories.
Using GitHub's search feature, we identified relevant PRs with the queries `\texttt{repo:\{repo\_name\} reviewed-by:github-actions[bot] is:pr}' for inline comments and `\texttt{repo:\{repo\_name\} commenter:github-actions is:pr}' for general comments.
Subsequently, we retrieved specific comments posted by \texttt{github-actions[bot]} (the account associated with comments generated by a GitHub Action) via the REST API endpoints: `\texttt{pulls/\{pull\_number\}/comments}' for inline comments and `\texttt{issues/\{pull\_number\}/comments}' for general comments.

\begin{table}[t]
\scriptsize
\centering
\begin{revision}
\caption{\textbf{RQ1:} \textbf{Adoption statistics} of 16 AI-based review actions, grouped by PR/File/Hunk review actions. ``Total/Mature Repos'' counts repositories configuring the action (Mature: $\ge$50 PRs). The observed activity columns detail the adoption distribution across repos, PRs, and comments in mature repos. Asterisk (*) indicates general instead of inline comments.}
\label{tab:adoption_overview}
\setlength{\tabcolsep}{6pt}
\resizebox{0.43\textwidth}{!}{
\begin{tabular}{c|c|c|ccc}
\hline
\multirow{2}{*}{\textbf{ID}} 
& \multirow{2}{*}{\makecell{\textbf{Total} \\ \textbf{Repos}}} 
& \multirow{2}{*}{\makecell{\textbf{Mature} \\ \textbf{Repos}}}
& \multicolumn{3}{c}{\textbf{Observed Review Activity}} \\
\cline{4-6}
& & & \textbf{Repos} & \textbf{PRs} & \textbf{Comments} \\
\hline
\multicolumn{6}{l}{\textbf{PR-level review actions: }} \\
\hline
5  & 7  & 2  & 0 & 0 & 0 \\
6  & 4  & 1  & 0 & 0 & 0 \\
8  & 3  & 1  & 1 & 24 & 37* \\
9  & 5  & 1  & 0 & 0 & 0 \\
11 & 8  & 4  & 1 & 38 & 38* \\
16 & 13 & 5  & 5 & 74 & 133* \\
\textbf{Total} & \textbf{40} & \textbf{14} & \textbf{7} & \textbf{136} & \textbf{208} \\
\hline
\multicolumn{6}{l}{\textbf{File-level review actions: }} \\
\hline
1  & 460 & 114 & 74 & 2,238 & 19,549 \\
2  & 37  & 11  & 9 & 421 & 421* \\
13 & 3   & 1   & 0 & 0 & 0 \\
15 & 4   & 1   & 1 & 1 & 3 \\
\textbf{Total} & \textbf{504} & \textbf{127} & \textbf{84} & \textbf{2,660} & \textbf{19,973} \\
\hline
\multicolumn{6}{l}{\textbf{Hunk-level review actions: }} \\
\hline
3  & 108 & 27 & 14 & 171 & 1,911 \\
4  & 47  & 8  & 5 & 28 & 193 \\
7  & 5   & 0  & 0 & 0 & 0 \\
10 & 6   & 1  & 1 & 4 & 31 \\
12 & 6   & 1  & 1 & 3 & 10 \\
14 & 2   & 0  & 0 & 0 & 0 \\
\textbf{Total} & \textbf{174} & \textbf{37} & \textbf{21} & \textbf{206} & \textbf{2,145} \\
\hline
\textbf{All 16} & \textbf{718} & \textbf{178} & \textbf{112} & \textbf{3,002} & \textbf{22,326} \\
\hline
\end{tabular}}
\end{revision}
\end{table}

Beyond adoption, we investigated how developers configure these actions, particularly focusing on optional parameters that reflect developer intent beyond mandatory execution requirements, and how these configurations evolve.
We extracted optional parameters from each action's documentation, noting their considerable diversity (in both number and function).
For example, \texttt{coderabbitai/ai-pr-reviewer} exposes 20 optional parameters to support its complex review workflows, whereas \texttt{presubmit/ai-reviewer} only defines three mandatory parameters.
To facilitate comparison, the first author manually grouped all optional parameters into nine functional categories, summarized in Table~\ref{tab:config_categories}. This grouping was reviewed by one co-author.
For each repository, we considered the workflow configuration file at its latest commit as the final configuration and used \texttt{PyYAML} to parse configurations according to our nine categories. 
To explore the evolution of the configuration, we calculated the number of commits and the time interval between the initial commit (introducing the action) and the latest commit.
If modifications occurred, we further compared specific settings in the initial and final configurations to identify frequently adjusted parameters.

\begin{table*}[t]
\scriptsize
\centering
\caption{\textbf{RQ2:} Refined Dataset for \textbf{Comment Addressing} Analysis (N=5,652): Post-review File Change Distribution by Comment Sources.}
\label{tab:file_changes_after_review}
\resizebox{0.96\textwidth}{!}{
\begin{tabular}{c|c|c|ccccc}
\hline
\textbf{ID} & \textbf{Action} & \textbf{Total Comments} & \textbf{Modified} & \textbf{Renamed-Modified} & \textbf{Renamed-Only} & \textbf{Deleted} & \textbf{Unchanged} \\
\hline
1 & anc95/ChatGPT-CodeReview & 2,831 & 436 (15.40\%) & 9 (0.32\%) & 0 (0.00\%) & 2 (0.07\%) & 2,384 (84.21\%) \\
2 & mattzcarey/code-review-gpt & 773 & 20 (2.59\%) & 0 (0.00\%) & 0 (0.00\%) & 0 (0.00\%) & 753 (97.41\%) \\
- & \textbf{File-level Review Actions Total} & \textbf{3,604} & \textbf{456 (13.89\%)} & \textbf{9 (0.27\%)} & \textbf{0 (0.00\%)} & \textbf{2 (0.06\%)} & \textbf{3,137 (85.78\%)} \\
\hline
3 & coderabbitai/ai-pr-reviewer & 713 & 303 (42.50\%) & 21 (2.95\%) & 3 (0.42\%) & 10 (1.40\%) & 376 (52.73\%) \\
4 & aidar-freeed/ai-codereviewer & 169 & 84 (49.70\%) & 0 (0.00\%) & 0 (0.00\%) & 3 (1.78\%) & 82 (48.52\%) \\
- & \textbf{Hunk-level Review Actions Total} & \textbf{882} & \textbf{387 (43.88\%)} & \textbf{21 (2.38\%)} & \textbf{3 (0.34\%)} & \textbf{13 (1.47\%)} & \textbf{458 (51.93\%)} \\
\hline
- & \textbf{Human Review Total} & \textbf{1,166} & \textbf{988 (84.75\%)} & \textbf{44 (3.77\%)} & \textbf{0 (0.00\%)} & \textbf{38 (3.26\%)} & \textbf{96 (8.23\%)} \\
\hline
\end{tabular}}
\vspace{-3pt}
\end{table*}


\vspace{1ex}
\noindent
\textbf{Results:}
Table~\ref{tab:adoption_overview} summarizes the usage of the 16 AI-based code review actions.
Out of 718 matched repositories, 178 met the maturity criterion ($\geq\!50$ PRs), and these contained a total of 22,326 AI-generated review comments.
We found that 37.1\% of the mature repositories declared an action but showed no generated comments, indicating a gap between declaration and actual use.
Among the remaining repositories, usage was highly concentrated on the four most popular actions. Actions ID-1 to ID-4 (including two file-level and two hunk-level reviewers) accounted for 91.1\% of the reviewed repositories, 95.2\% of the pull requests, and 98.9\% of the generated comments.
This was largely driven by the widespread use of ID-1 \texttt{anc95/ChatGPT-CodeReview}.
In contrast, PR-level review actions saw limited adoption and impact.

Regarding configuration evolution, the majority of the mature repositories (147/178, 82.6\%) customized at least one optional parameter.
The most frequently configured options were \textit{Prompt Context Augmentation} (103/147, 70.1\%), primarily for specifying the natural language of review comments, \textit{LLM Selection} (95, 64.6\%), and \textit{Input Settings} (70, 47.6\%).
On average, these 178 repositories made changes to the configuration file 4 times after the initial setup.
Of these, 68 repositories (38.2\%) retained their original configurations without further changes. Temporal analysis revealed that 50 projects (28.1\%) finalized modifications within one week, and 20 (11.2\%) within one month, while 40 repositories (22.5\%) continued adjustments beyond one month.
The most frequently modified parameters were \textit{LLM Selection} (25, 22.7\% of 110 repositories that made post-setup changes), \textit{Prompt Customization} (21, 19.1\%), and \textit{LLM Hyperparameters} (13, 11.8\%).
Beyond action-specific options, we also observe that 50 repositories (45.5\%) adjusted their workflow-level triggers by modifying the \texttt{on} or \texttt{if} conditions.
Additionally, 12 repositories (10.9\%) updated the action reference, such as replacing \texttt{@main} with a specific release tag.

\begin{rqanswerbox}
\textbf{Answer to RQ1:} We analyzed \textbf{178} mature repositories and found a total of \textbf{22,326} AI-generated review comments. Among the 178 repositories, \textbf{82.6\%} customized at least one optional parameter, showing that developers often go beyond default settings when using these actions.
\end{rqanswerbox}

\begin{figure}[t] 
	\centering
     \begin{revision}
     \setlength{\abovecaptionskip}{3pt}
	\includegraphics[width=0.49\textwidth]{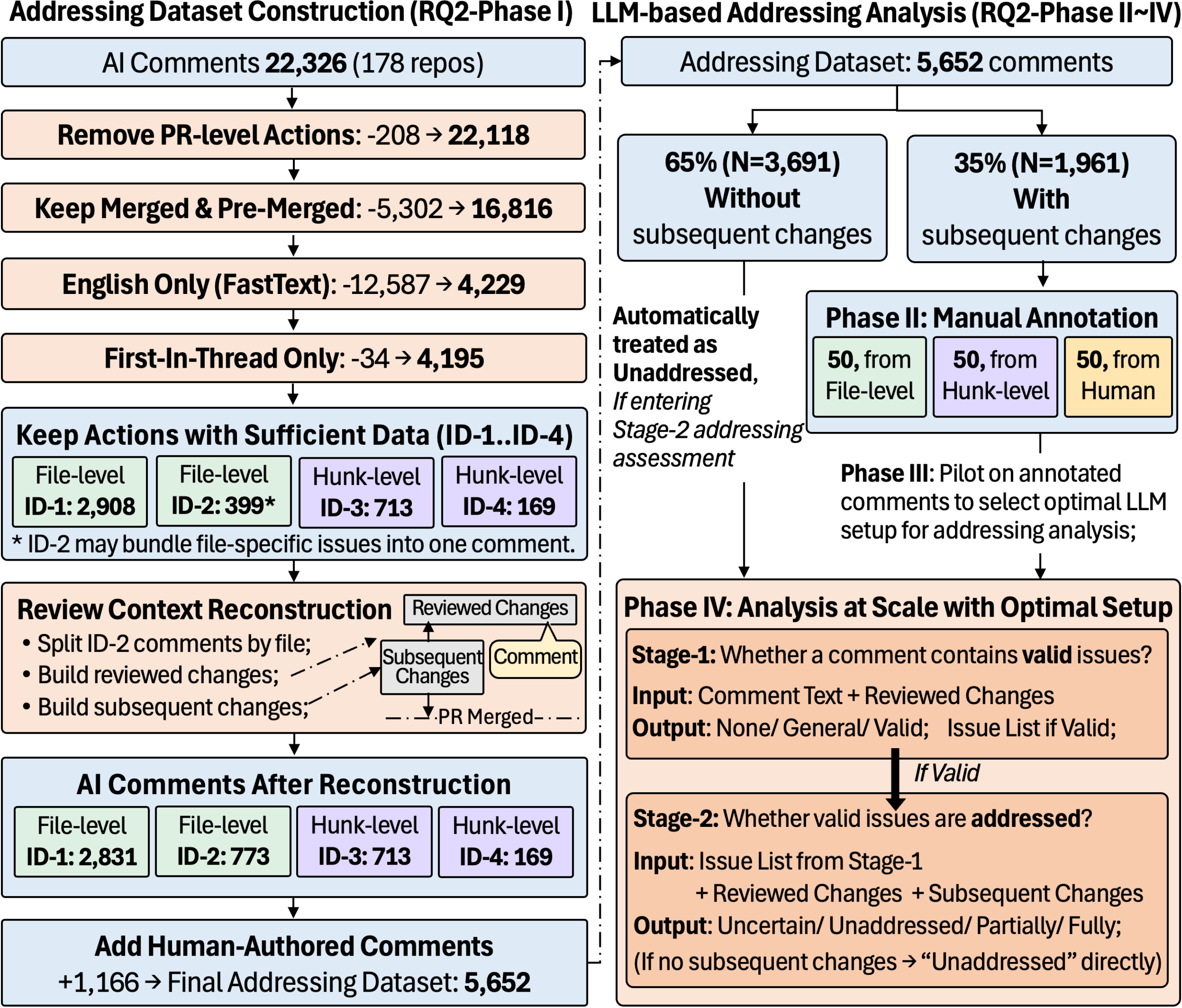} 
	\caption{\textbf{Overview} of the \textbf{RQ2} comment addressing analysis pipeline.}
	\label{fig:RQ2_pipeline}
    \end{revision}
\end{figure}

\begin{revision}
\subsection{\textbf{RQ2:} To what extent do AI-generated review comments lead to code changes compared to human review comments?}

\vspace{1ex}
\noindent
\textbf{Approach:}
To assess the practical impact of these actions, we leverage LLMs to extensively examine whether the generated review comments led to actual code changes (which we define as ``addressed'').
We use the term ``comment addressing analysis'' to refer to our analysis of whether comments have been addressed.
We first created the ``comment addressing dataset'' (filtered from \texttt{RQ1} and augmented with human-written review comments) to carry out this analysis. 
Next, we sampled a subset to manually annotate whether they had been addressed and evaluated LLM performance on this subset. 
Finally, we employed the best-performing LLM approach across the entire dataset.
Figure~\ref{fig:RQ2_pipeline} presents an overview of this workflow.
\end{revision}

\vspace{1ex}
\noindent
\textbf{\textit{Phase I: Refining the Subset for Comment Addressing Analysis.}}
We first excluded comments from PR-level actions because of their significant differences from typical human code review practices and their low volume.
For comments from file-level and hunk-level actions, we only retained those in merged PRs (and prior to merging) to capture finalized developer decisions, yielding 16,816 comments.
To facilitate human analysis, we selected only English comments using FastText, reducing the dataset to 4,229 comments (a removal of 12,587 comments).
ID-1 (\texttt{anc95/ChatGPT-CodeReview}) saw a significant reduction in comments at this stage, with 11,744 comments (mostly in Korean) filtered out.
We further discuss potential threats introduced by this language-based filtering in Section~\ref{sec:threats}.
Given that some actions enable user-LLM dialogue within threads, we restricted our analysis to first-in-thread comments to focus on the original review suggestions rather than follow-up interactions (excluding an additional 34 comments).
After these filters, only Actions ID-1 to ID-4 retained sufficient data for the addressing analysis, with 2,908, 399, 713, and 169 comments respectively.

Next, we constructed the necessary context for assessing comment addressing, which included the originally reviewed changes and subsequent code changes within the reviewed file.
For hunk-level reviews, which were all submitted as inline comments, GitHub’s API provides sufficient metadata to reconstruct the reviewed code.
Specifically, \texttt{original\_commit\_id} identifies the commit under review, \texttt{original\_start\_line} and \texttt{original\_line} specify the range of reviewed lines, and \texttt{diff\_hunk} captures the change block where the comment was made. 
We extracted the reviewed code by slicing \texttt{diff\_hunk} based on the recorded line range. 
For single-line comments (with only \texttt{original\_line}), we additionally included the three preceding lines to align with GitHub's display convention for diffs.
File-level actions required additional handling, as they often cover multiple changes within a single comment.
For Action ID-1 (\texttt{anc95/ChatGPT-CodeReview}), we simulated its two strategies to define the reviewed changes: (1) directly extracting the file's changes from the \texttt{original\_commit\_id},
and (2) computing cumulative diffs between the PR's base commit and \texttt{original\_commit\_id} via GitHub’s ``\texttt{/compare}'' endpoint.
We retained only those comments whose \texttt{diff\_hunk} could match as a subset of either reconstructed change set, and discarded inconsistent cases potentially caused by temporal shifts in the PR's base commit.
For Action ID-2 (\texttt{mattzcarey/code-review-gpt}), which posts general comments (without \texttt{original\_commit\_id}) reviewing multiple files, we first split comments by file, then approximated the reviewed commit as the one immediately preceding the comment’s creation.
We reconstructed the reviewed changes by comparing this commit against the PR’s base commit.

For subsequent changes within reviewed files, we defined these as cumulative changes between the reviewed and merged versions.
We retrieved file content at \texttt{original\_commit\_id} and at the merge commit using ``\texttt{contents/\{file\_path\}?ref=\{sha\}}'' endpoint, then calculated their textual differences using the \texttt{difflib} Python library.
We chose this approach over the ``\texttt{/compare}'' endpoint to avoid exaggerating rebase-related changes.
We also accounted for file renaming by resolving to their final paths.
Subsequent changes were categorized as follows:
(1) \textbf{\textit{Modified}} (content changed under original filename);
(2) \textbf{\textit{Renamed-Modified}} (filename changed and content modified);
(3) \textbf{\textit{Renamed-Only}} (filename changed with no content modified);
(4) \textbf{\textit{Deleted}} (file removed);
(5) \textbf{\textit{Unchanged}} (no modifications or renaming occurred).
Table~\ref{tab:file_changes_after_review} shows the distributions of subsequent changes by comment source. 
In total, we obtained 4,486 action-generated comments from 51 repositories.

For comparison, we collected human-authored review comments from the same 51 repositories for which we had obtained action-generated comments in Phase I.
For each repository, we defined its AI review activation period as the time span between the first and last action-generated comment.
We restricted our collection of human-written comments to this same time window to minimize potential effects on the addressing ratio (e.g., differences in the project’s development phase).
Within this period, we collected human-authored inline comments that met the same filtering criteria as the action-generated comments:
(1) from merged pull requests and posted prior to merging,
(2) written in English,
and (3) first in their comment threads (this filter excluded nearly half of the candidate comments).
We also excluded comments from accounts whose login names contain the string ``bot''~\cite{golzadeh2020bot}.
This process yielded a final set of 1,166 human-authored review comments.
Following the same procedure used for hunk-level review actions, we reconstructed their originally reviewed changes and subsequent file modifications.
Table~\ref{tab:file_changes_after_review} summarizes the statistics for the human comments.
In total, our comment addressing dataset includes 5,652 review comments: 3,604 from file-level review actions, 882 from hunk-level actions, and 1,166 authored by human reviewers.

\vspace{1ex}
\noindent
\textbf{\textit{Phase II: Establishing Ground Truth via Human Annotation.}}
In this phase, we sampled a subset and manually evaluated whether they had been addressed to serve as the ground truth. Specifically, we randomly sampled 50 comments from each category where subsequent file modifications occurred (i.e. 467 eligible from file-level actions, 424 eligible  from hunk-level actions, and 1,070 eligible  human-authored comments), yielding 150 comments in total.
Before the full annotation, two authors first established an annotation guideline through consensus on 9 comments (the first 3 from each category).
Through this process, we identified two challenges that prevent comment addressing from being a simple binary classification.
First, some comments did not contain any suggestions, such as simple praise in human reviews (e.g., ``\textit{Amazing!}''), or mere code summaries in action-generated reviews.
Some comments contained overly general suggestions lacking clear guidance (e.g., ``\textit{Suggest to test thoroughly}''), making it difficult to determine whether they were addressed.
Second, a single comment may include multiple issues or suggestions, requiring a more granular assessment of their resolution.
Therefore, we designed a two-stage annotation task.
In \textbf{Stage-1}, we examined the comment text and the reviewed changes to determine whether a comment contained valid (i.e. specific and actionable) issues or suggestions, resulting in the following classifications:
\begin{itemize}[leftmargin=1.5em]
\item \textbf{\textit{None}}: Does not contain any issues/suggestions.
\item \textbf{\textit{General}}: Only contains overly general issues/suggestions.
\item \textbf{\textit{Valid}}: Contains at least one valid issues/suggestions.
\end{itemize}
In \textbf{Stage-2}, for the comments classified as ``Valid'' in Stage-1, we further analyzed subsequent code changes to determine the extent to which the identified issues or suggestions were addressed.
The possible outcomes were:
\begin{itemize}[leftmargin=1.5em]
\item \textbf{\textit{Valid-Uncertain}}: Insufficient information to determine (mainly caused by file deletion or large-scale refactoring). 
\item \textbf{\textit{Valid-Unaddressed}}: None of identified items addressed.
\item \textbf{\textit{Valid-Partially}}: Some but not all items addressed (at least one of them was completely addressed).
\item \textbf{\textit{Valid-Fully}}: All items addressed.
\end{itemize}
Following this guideline, the remaining 141 comments (47 per category) were independently labeled by one author and an external graduate student in software engineering who is not a co-author of this paper.
Inter-rater agreement (Cohen's $\kappa$) reached substantial levels over the full 6-class annotation scheme (i.e., ``None'', ``General'', ``Valid-Uncertain'', ``Valid-Unaddressed'', ``Valid-Partially'', ``Valid-Fully''): 0.674 for file-level actions, 0.734 for hunk-level actions, and 0.764 for human reviews.
A third author resolved disagreements by selecting between two annotations or making a new judgment when necessary.
Table~\ref{fig:annotation_results} shows the final label distribution across 150 annotated comments.

\vspace{1ex}
\noindent
\textbf{\textit{Phase III: Evaluating LLM Effectiveness for Comment Addressing Analysis.}}
Based on our human-annotated dataset, we evaluated the performance of three families of LLMs for addressing analysis:
OpenAI~\cite{OpenAI} (\textit{gpt-4.1}, \textit{gpt-4o}, and \textit{o4-mini} / \textit{o3-mini} with medium reasoning effort), 
Claude~\cite{Claude} (\textit{claude-3-sonnet} and \textit{claude-3-haiku}), 
and DeepSeek~\cite{DeepSeek} (\textit{deepseek-r1} and \textit{deepseek-v3}).
The specific API endpoints and model versions used are documented in our online appendix scripts; the evaluations were performed in April and May 2025.
Following the human annotation process, we designed separate prompts for Stage-1 and Stage-2.
Stage-1 aimed to classify comments as ``None'', ``General'', or ``Valid'', and extract a list of valid issues or suggestions for ``Valid'' comments. 
When Stage-1 identified a comment as ``Valid'', Stage-2 was triggered to determine its final addressed status.
Both stages utilized the same contextual information (i.e. reviewed changes and subsequent file modifications) and the classification scheme as the human annotators.
The details of the LLM-assisted framework with specific prompts are available in the online appendix for other researchers to use.

\begin{table}[t]
\scriptsize
\centering
\caption{\textbf{RQ2:} Annotation \textbf{Label Distribution} by Comment Sources.}
\label{fig:annotation_results}
\setlength{\tabcolsep}{2pt} 
\begin{adjustbox}{max width=0.48\textwidth}
\renewcommand{\arraystretch}{1.1}
\begin{tabular}{>{\bfseries}c|c|cc|cc|cc}
\hline
\multirow{2}[3]{*}{\textbf{Source}} & 
\multirow{2}[3]{*}{\textbf{Total}} & 
\multicolumn{2}{c|}{\textbf{Invalid}} & 
\multicolumn{2}{c|}{\textbf{Valid, Not Addressed}} & 
\multicolumn{2}{c}{\textbf{Valid, Addressed}} \\
\cline{3-8}
& & \multicolumn{1}{>{\bfseries\itshape}c}{None} & 
\multicolumn{1}{>{\bfseries\itshape}c|}{General} & 
\multicolumn{1}{>{\bfseries\itshape}c}{\makecell{Valid-\\Uncertain}} & 
\multicolumn{1}{>{\bfseries\itshape}c|}{\makecell{Valid-\\Unaddressed}} & 
\multicolumn{1}{>{\bfseries\itshape}c}{\makecell{Valid-\\Partially}} & 
\multicolumn{1}{>{\bfseries\itshape}c}{\makecell{Valid-\\Fully}} \\
\hline

\multirow{2}{*}{\makecell{File-level\\Action}} 
& \multirow{2}{*}{50} 
& 2.0\% & 30.0\% & 0.0\% & 42.0\% & 22.0\% & 4.0\% \\
\cline{3-8}
& & \multicolumn{2}{c|}{32.0\%} & \multicolumn{2}{c|}{42.0\%} & \multicolumn{2}{c}{26.0\%} \\
\cline{3-8}
\hline

\multirow{2}{*}{\makecell{Hunk-level\\Action}} 
& \multirow{2}{*}{50} 
& 8.0\% & 8.0\% & 6.0\% & 42.0\% & 2.0\% & 34.0\% \\
\cline{3-8}
& & \multicolumn{2}{c|}{16.0\%} & \multicolumn{2}{c|}{48.0\%} & \multicolumn{2}{c}{36.0\%} \\
\cline{3-8}
\hline

\multirow{2}{*}{Human} 
& \multirow{2}{*}{50} 
& 12\% & 0.0\% & 6.0\% & 2.0\% & 2.0\% & 78.0\% \\
\cline{3-8}
& & \multicolumn{2}{c|}{12.0\%} & \multicolumn{2}{c|}{8.0\%} & \multicolumn{2}{c}{80.0\%} \\
\hline
\end{tabular}
\end{adjustbox}

\vspace{2em}

\begin{revision}
\scriptsize
\centering
\caption{
\textbf{RQ2:}
\textbf{Performance} on \textbf{150 annotated comments} using the best automated setup, with \textit{openai-gpt-4.1} for Stage-1 and \textit{openai-o3-mini} for Stage-2.
The last column reports the ability to distinguish ``Valid, Addressed'' comments from all others.
``Avg.'': average across three sources.}
\label{tab:llm_eval_metrics}
\setlength{\tabcolsep}{2pt}
\renewcommand{\arraystretch}{1.1}
\begin{adjustbox}{max width=0.49\textwidth}
\begin{tabular}{c|cc|cc|cc|cc}
\hline
\multirow{2}[1]{*}{\textbf{Source}}  & 
\multicolumn{2}{c|}{\textbf{Full 6-class}} & 
\multicolumn{2}{c|}{\makecell{\textbf{Stage-1}\\\textbf{(Validity)}}} & 
\multicolumn{2}{c|}{\makecell{\textbf{Stage-2}\\\textbf{(Addressing)}}} & 
\multicolumn{2}{c}{\makecell{\textit{\textbf{Valid, Addressed}}\\\textit{\textbf{vs. All Others}}}} \\
\cline{2-9}
& \textbf{OA} & \textbf{Macro-F1} 
& \textbf{OA} & \textbf{Macro-F1} 
& \textbf{OA} & \textbf{Macro-F1} 
& \textbf{OA} & \textbf{Macro-F1} \\
\hline
\makecell{\textbf{File-level}\\\textbf{Action}}   
& 83.6\% & 78.1\% 
& 94.4\% & 93.5\% 
& 88.5\% & 88.4\% 
& 92.0\% & 90.5\%
\\ \hline
\makecell{\textbf{Hunk-level}\\\textbf{Action}}   
& 89.2\% & 83.4\% 
& 95.2\% & 90.8\% 
& 93.2\% & 93.1\% 
& 91.6\% & 91.2\%
\\ \hline
\textbf{Human}              
& 85.6\% & 62.4\% 
& 94.0\% & 88.0\% 
& 93.7\% & 83.9\% 
& 90.8\% & 87.6\%
\\ \hline
\textbf{Avg.}    
& \textbf{86.1\%} & \textbf{74.6\%} 
& \textbf{94.5\%} & \textbf{90.8\%} 
& \textbf{91.8\%} & \textbf{88.5\%} 
& \textbf{91.5\%} & \textbf{89.8\%} \\
\hline
\end{tabular}
\end{adjustbox}

\vspace{2em}

\caption{\textbf{RQ2:}
Performance of \textbf{distinguishing ``Valid, Addressed'' comments from others} across the \textbf{entire dataset}. Unchanged Group is logically classified as non-addressed (100\% Acc.). }
\label{tab:addressing_accuracy_breakdown}
\centering
\scriptsize
\setlength{\tabcolsep}{2.5pt}
\begin{tabular}{c| >{\centering\arraybackslash}p{0.9cm} >{\centering\arraybackslash}p{0.9cm} | >{\centering\arraybackslash}p{0.9cm} >{\centering\arraybackslash}p{0.9cm} | >{\centering\arraybackslash}p{0.74cm} >{\centering\arraybackslash}p{0.74cm}}
\hline
\multirow{2}{*}{\textbf{Source}} & \multicolumn{2}{c|}{\textbf{No Code Change}} & \multicolumn{2}{c|}{\textbf{With Code Change}} & \multicolumn{2}{c}{\textbf{Overall}} \\
 & \textbf{Total} & \textbf{Acc.} & \textbf{Total} & \textbf{Acc.} & \textbf{Total} & \textbf{Acc.} \\ \hline
\textbf{File-level Action} & 3,137 & 100\% & 467 & 92.0\% & 3,604 & 99.0\% \\ 
\textbf{Hunk-level Action} & 458 & 100\% & 424 & 91.6\% & 882 & 96.0\% \\
\textbf{Human} & 96 & 100\% & 1,070 & 90.8\% & 1,166 & 91.6\% \\
\hline
\textbf{Total} & \textbf{3,691} & \textbf{100\%} & \textbf{1,961} & \textbf{91.3\%} & \textbf{5,652} & \textbf{97.0\%} \\ \hline
\end{tabular}
\vspace{-3pt}
\end{revision}

\end{table}

We decoupled the stages to allow different model selection based on stage-specific performance.
With temperature set to 0, we ran each model five times for robust evaluation.
Performance was measured using overall accuracy (OA) and Macro-F1 scores. 
In the first round of evaluation, we used the same LLM for both Stage 1 and Stage 2.
Beyond assessing performance on the final 6-class scheme, we also evaluated the models' effectiveness in each stage independently.
For Stage-1 (detecting valid comment), we grouped ``None'' and ``General'' into a ``Not Valid'' category, and reported the models' ability to distinguish between ``Not Valid'' and ``Valid'' comments.
For Stage-2 (addressed assessment), focusing on comments correctly identified as ``Valid'', we grouped ``Valid-Uncertain'', ``Valid-Unaddressed'' into ``Valid, Not Addressed,'' and ``Valid-Partially'' and ``Valid-Fully'' into ``Valid, Addressed'', then reported the models' ability to differentiate these two outcomes.
In the second round of evaluation, we cross-combined the top three models from each stage to find the optimal combination.

\begin{table}[t]
\scriptsize
\centering
\caption{\textbf{RQ2:} LLM-Assigned \textbf{Addressing Labels} for Comments by Source (ID-1: anc95/ChatGPT-CodeReview, ID-2: mattzcarey /code-review-gpt, ID-3: coderabbitai/ai-pr-reviewer, ID-4: aidar-freeed/ai-codereviewer, Human Reviews).
Distribution shift compared with Table~\ref{fig:annotation_results} results from a different sampling strategy (see RQ2's results).}
\label{fig:annotation_results_detail}
\setlength{\tabcolsep}{2pt}
\begin{adjustbox}{max width=0.48\textwidth}
\renewcommand{\arraystretch}{1.1}
\begin{tabular}{>{\bfseries}c@{}|c|cc|cc|cc}
\hline
\multirow{2}[3]{*}{\textbf{Source}} & 
\multirow{2}[3]{*}{\textbf{Total}} & 
\multicolumn{2}{c|}{\textbf{Invalid}} & 
\multicolumn{2}{c|}{\textbf{Valid, Not Addressed}} & 
\multicolumn{2}{c}{\textbf{Valid, Addressed}} \\
\cline{3-8}
& & \multicolumn{1}{>{\bfseries\itshape}c}{None} & 
\multicolumn{1}{>{\bfseries\itshape}c|}{General} & 
\multicolumn{1}{>{\bfseries\itshape}c}{\makecell{Valid-\\Uncertain}} & 
\multicolumn{1}{>{\bfseries\itshape}c|}{\makecell{Valid-\\Unaddressed}} & 
\multicolumn{1}{>{\bfseries\itshape}c}{\makecell{Valid-\\Partially}} & 
\multicolumn{1}{>{\bfseries\itshape}c}{\makecell{Valid-\\Fully}} \\
\hline

\multicolumn{8}{l}{\textbf{File-level Review Action:}} \\
\hline
\multirow{2}{*}{ID-1} & \multirow{2}{*}{2,831} & 15.7\% & 24.8\% & 0.2\% & 55.2\% & 3.7\% & 0.5\% \\
            \cline{3-8}
            &      & \multicolumn{2}{c|}{40.4\%} & \multicolumn{2}{c|}{55.4\%} & \multicolumn{2}{c}{4.2\%} \\
\hline
\multirow{2}{*}{ID-2} & \multirow{2}{*}{773}  & 1.6\%  & 18.1\% & 0.1\% & 79.3\% & 0.8\% & 0.1\% \\
            \cline{3-8}
            &      & \multicolumn{2}{c|}{19.7\%} & \multicolumn{2}{c|}{79.4\%} & \multicolumn{2}{c}{0.9\%} \\
\hline
\hline
\multicolumn{8}{l}{\textbf{Hunk-level Review Action:}} \\
\hline
\multirow{2}{*}{ID-3} & \multirow{2}{*}{713}  & 15.4\% & 7.4\%  & 2.2\% & 55.7\% & 4.2\% & 15.0\% \\
            \cline{3-8}
            &      & \multicolumn{2}{c|}{22.9\%} & \multicolumn{2}{c|}{57.9\%} & \multicolumn{2}{c}{19.2\%} \\
\hline
\multirow{2}{*}{ID-4} & \multirow{2}{*}{169}  & 3.6\%  & 2.4\%  & 2.4\% & 85.2\% & 0.6\% & 5.9\% \\
            \cline{3-8}
            &      & \multicolumn{2}{c|}{5.9\%}  & \multicolumn{2}{c|}{87.6\%} & \multicolumn{2}{c}{6.5\%} \\
\hline
\hline
\multirow{2}{*}{Human}       & \multirow{2}{*}{1,166} & 15.9\% & 3.7\%  & 4.2\% & 16.3\% & 4.0\% & 56.0\% \\
            \cline{3-8}
            &      & \multicolumn{2}{c|}{19.6\%} & \multicolumn{2}{c|}{20.5\%} & \multicolumn{2}{c}{60.0\%} \\
\hline
\end{tabular}
\end{adjustbox}
\vspace{-3pt}
\end{table}

\begin{revision}
\vspace{1ex}
\noindent
\textbf{\textit{Phase IV: LLM-Based Labeling of Comment Addressing.}} 
Finally, we apply the best-performing setup determined in Phase III to conduct comment addressing analysis on the entire dataset of 5,652 review comments.
\end{revision}

\vspace{1ex}
\noindent
\textbf{Results:}
We first present the evaluation results from Phase III.
In our first round of evaluation, we observed that general-purpose models usually performed well in Stage-1, while reasoning-enhanced models excelled in Stage-2.
The top performers for Stage-1 were \textit{gpt-4.1} (94.5\% average overall accuracy), \textit{deepseek-v3} (94.0\%), and \textit{claude-3-sonnet} (93.2\%).
For Stage-2, \textit{deepseek-r1} led with 95.4\%, followed by \textit{openai-o3-mini} (92.4\%) and \textit{o4-mini} (91.3\%).
\begin{revision}
Table~\ref{tab:llm_eval_metrics} reports the performance of our optimal cross-combined setup: \textit{gpt-4.1} for Stage-1 and \textit{o3-mini} for Stage-2, which achieved strong and balanced performances across three comment sources, with an average of 86.1\% overall accuracy and 74.6\% Macro-F1 score under the full 6-class scheme.
The lower macro-F1 for human-authored comments (62.4\%) reflects stronger class imbalance within that subset (see Table~\ref{fig:annotation_results}), which penalizes macro-averaged metrics.
Importantly, performance on the primary downstream distinction (``Valid, Addressed'' vs. all others) remains high, with an average of 91.5\% overall accuracy and 89.8\% macro-F1.
Furthermore, when analyzing the complete dataset, we found that a significant portion of comments (87.0\% of file-level, 51.9\% of hunk-level, and 8.2\% of human comments) had no subsequent modifications.
We can thus trivially classify them as non-``Valid, Addressed'' with certainty.
Considering the natural class imbalance in real-world review data, the overall classification accuracy across the full dataset is 97\%, ensuring the reliability of our subsequent large-scale analysis in Phase IV (detailed breakdown in Table~\ref{tab:addressing_accuracy_breakdown}).
\end{revision}


Following the best-performing setup on the annotated dataset, we used \textit{gpt-4.1} to assess the validity of all 5,652 comments, identifying 3,955 as valid, and then used \textit{openai-o3-mini} to assess whether these valid comments had been addressed.
Table~\ref{fig:annotation_results_detail} details the results across the entire dataset.
Comparing Table~\ref{fig:annotation_results} (manual labels) and Table~\ref{fig:annotation_results_detail} (automated labels), we observe a discrepancy between their annotation distributions.
This stems from our sampling strategy.
For the manual annotation, we purposefully sampled only those comments where the reviewed file was subsequently modified.
This was a necessary precondition for assessing if a comment was addressed, but it naturally biased the sample toward ``Valid'' and ``Addressed'' classifications.
In contrast, the automated annotation was applied to the entire comment dataset, which included a large volume of comments (65.3\%) with no subsequent file changes.
Following our procedure, if these comments were identified as ``Valid'' in Stage-1, they were automatically classified as ``Not Addressed'' in Stage-2.
Consequently, the proportion of ``Valid, Addressed'' comments is lower in the automated setting.
\begin{revision}
Beyond this discrepancy, both tables indicate that the addressing rate of AI-generated review comments (0.9\%--19.2\% in Table~\ref{fig:annotation_results_detail}) still lags behind human review comments (60\%).
Overall, hunk-level review actions exhibit a higher addressing rate (6.5\%--19.2\%) compared to file-level actions (0.9\%--4.2\%).
\end{revision}

\begin{rqanswerbox}
\begin{revision}
\textbf{Answer to RQ2:}
We found that \textbf{hunk-level review actions (6.5\%--19.2\%)} exhibit a higher addressing rate compared to \textbf{file-level actions (0.9\%--4.2\%)}, yet the addressing rate of AI-generated review comments still lags behind \textbf{human review comments (60\%)}.
\end{revision}
\end{rqanswerbox}

\subsection{\textbf{RQ3:} Which factors impact the likelihood that code review comments lead to code changes?}
\vspace{1ex}
\noindent
\textbf{Approach:}
To answer RQ3, we conducted an interpretable analysis to identify the factors influencing whether review comments lead to code changes.
We focus specifically on the 3,879 valid comments from RQ2 that have a definitive addressing label (excluding 76 uncertain cases), examining the factors that determine whether they are addressed.
For invalid comments, we discuss their potential causes and mitigation strategies in Section~\ref{sec:discussion}.
To quantify the impact of different factors, we engineered a structured feature set covering various dimensions that potentially affect response behavior.
Since direct interpretation of LLM decision-making is challenging, we trained a Random Forest classifier to fit the addressing results derived in RQ2 and employed SHAP (SHapley Additive exPlanations)~\cite{DBLP:conf/nips/LundbergL17} analysis to interpret the influence and directionality of each feature on the model's predictions.

\begin{table}[t]
\centering
\begin{revision}
\caption{\textbf{RQ3}: \textbf{36 Selected Features} Across Four Dimensions Used for Analyzing Factors Impacting Comment Addressing.}
\label{tab:feature_table}
\setlength{\tabcolsep}{1pt}
\begin{adjustbox}{max width=0.49\textwidth}
\renewcommand{\arraystretch}{1.1}
\begin{tabular}{c|l}
\hline
\multicolumn{2}{l}{\textbf{Source Features (10): Who generated the comment and how it was generated}} \\
\hline
Is\_Human & 1 if written by a human; 0 if AI-generated \\
Is\_File\_Level\_Action & 1 if generated by file-level action; 0 otherwise \\
Is\_Action\_[4] & One-hot (4): Which action produced the comment \\
Trigger\_[auto/manual] & One-hot (2): Action trigger type: Auto/Manual (all-zero if human) \\
LLM\_[GPT-3.5/4] & One-hot (2): LLM used for reviewing (all-zero if human) \\
\hline
\multicolumn{2}{l}{\textbf{Repository Features (3): Repository-level context where the review occurred}} \\
\hline
Repo\_File\_Size & Total size of files in the repository (in bytes) \\
Repo\_Issue\_Count & Total number of issues in the repository \\
Repo\_Contributor\_Count & Number of non-bot contributors in the repository \\
\hline
\multicolumn{2}{l}{\textbf{Modification Features (12): Characteristics of the code changes under reviewed}} \\
\hline
Author\_Is\_Bot & 1 if the author of reviewed commit is a bot; 0 otherwise \\
Author\_Is\_Anon & 1 if the author is anonymous; 0 otherwise \\
Author\_Prior\_Commits & Number of previous commits by the author to this repository \\
Commit\_Del & Total lines deleted in changed files\\
Commit\_Base\_Lines & Total lines in all changed files at base commit \\
File\_Is\_Code\ & 1 if a certain reviewed file is a programming file (via Linguist) \\
File\_Depth & Directory nesting level of the file \\
File\_Add/Del & Lines added/deleted in the reviewed file\\
File\_Base\_Lines & Line count of the file at base commit \\
Comment\_Add/Del & Lines added/deleted the comment scope \\
\hline
\multicolumn{2}{l}{\textbf{Comment Features (11): Structure and content of the review comment itself}} \\
\hline
Prior\_Comment\_Len & Total character length of all prior comments in the PR \\
Inline\_Code & 1 if contains inline code snippets (\texttt{\`}code\texttt{\`}) \\
Multiline\_Code & 1 if contains multiline code blocks (\texttt{\`}\texttt{\`}\texttt{\`}code\texttt{\`}\texttt{\`}\texttt{\`}) \\
Text\_Length & Total Length of the current comment (in characters) \\
Code\_Text\_Ratio & Ratio of code length to total comment length \\
LDA\_Topic\_[6] & \makecell[l]{Six LDA topic probabilities with GPT-4.1 labeled themes: \\
0: Dependency Version Update and Compatibility; \\
1: UI Usability and Accessibility Enhancements; \\
2: Concurrency Control and Locking Robustness; \\
3: Module Import Optimization; \\
4: Variable Naming and Code Readability; \\
5: Error Handling and Code Documentation;} \\
\hline
\end{tabular}
\end{adjustbox}
\vspace{-3pt}
\end{revision}
\end{table}

\vspace{1ex}
\noindent
\textbf{\textit{Phase I: Structured Feature Engineering.}}
Inspired by previous works~\cite{rahman2017predicting,DBLP:conf/esem/DeyM20}, We initially built a 45-feature set across four dimensions to comprehensively capture the possible impacting factors.
\rev{After checking for multicollinearity, we removed 9 redundant features, yielding a final set of 36 features presented in Table~\ref{tab:feature_table}.}
The complete list of candidate features can be found in our online appendix.
Specifically, \textbf{Source Features} capture who generated the comment (a human developer or one of four actions) and two key aspects of action configuration: the trigger mode (automatic vs. manual) and the LLM used (GPT-3.5 vs. GPT-4).
These configurations were extracted by parsing the workflow file at the comment’s associated commit.
If the workflow included an `\texttt{if}' conditional trigger, we labeled the comment as manually triggered; otherwise, it was considered automatic.
\begin{revision}
To determine the LLM used for comment generation, we first checked for explicitly specified models in the workflow configuration. If unspecified ($n$=1,043), we manually inferred the default model by examining 9 associated action versions.
Specifically, 7 were fixed versions (e.g., \texttt{@v1}) with constant defaults, while 2 were rolling \texttt{main} branch; for the latter, we identified the default switch commit (e.g., GPT-3.5$\rightarrow$GPT-4) and labeled comments pre-/post-switch accordingly.
All four actions in our final analysis employed OpenAI models, which we mapped to either the GPT-3.5 or GPT-4 family.
\end{revision}
\textbf{Repository Features} reflect the project context, including overall size and activity level (e.g., number of issues).
\textbf{Modification Features} characterize the code changes under review, including the authorship and change properties.
We used \textit{Author\_Prior\_Commits} to estimate the author’s familiarity with the project.
Edge cases, such as when the author was identified as a bot or an anonymous user (1.8\% of valid comments), were flagged using \textit{Author\_Is\_Bot/Anon}, with \textit{Author\_Prior\_Commits} set to 0 accordingly.
We further described change scope at the commit, file, and comment levels, along with the volume of the involved files before changes were made.
\textbf{Comment Features} reflect the characteristics of the comment itself, including structural features (e.g., its position in the PR discussion thread) and textual features (e.g., code ratio).
To capture comment intent, we applied Latent Dirichlet Allocation (LDA)~\cite{blei2003latent} to the 3,879 comments.
Following hyperparameter tuning strategies in prior work~\cite{DBLP:conf/msr/Treude019}, we selected six as the optimal number of topics.
We then used \textit{gpt-4.1} to summarize the top ten comments for each topic to generate interpretable labels (as shown in Table~\ref{tab:feature_table}).
Finally, we applied AutoSpearman correlation analysis~\cite{DBLP:conf/icsm/JiarpakdeeTT18} ($\rho\!>\!0.7$ threshold) to address potential multicollinearity among features, and removed 9 redundant features (e.g., \textit{Repo\_File\_Count}, \textit{Commit\_Changed\_File\_Count}).

\begin{table}[t]
\scriptsize
\centering
\begin{revision}
\caption{\textbf{RQ3:} \textbf{top 10 features} influencing \textbf{total comment addressing}. Gray-shaded rows denote features that also rank among the top 10 for AI comment addressing (Table~\ref{tab:shap_feature_contribution_ai}).}
\label{tab:shap_feature_contribution_global}
\setlength{\tabcolsep}{3pt}
\begin{adjustbox}{max width=0.48\textwidth}
\renewcommand{\arraystretch}{1.1}
\begin{tabular}{c|c|c|c}
\hline
\textbf{Feature} & \textbf{Rank} & \textbf{Importance ($|\phi|$)} & \textbf{Directionality ($\rho$)} \\
\hline
\multicolumn{4}{l}{\textbf{Source Feature (10):} $\sum|\phi|$=0.1589; $\mu|\phi|$=0.0159} \\
\hline
Is\_Human & 1 & 0.0449 & 0.99 \\
Trigger\_auto & 2 & 0.0410 & -0.96 \\
\rowcolor{gray!20}Is\_File\_Level\_Action & 3 & 0.0346 & -0.95 \\
\hline
\multicolumn{4}{l}{\textbf{Repository Feature (3):} $\sum|\phi|$=0.0342; $\mu|\phi|$=0.0114} \\
\hline
\rowcolor{gray!20}Repo\_File\_Size & 5 & 0.0184 & 0.70 \\
\hline
\multicolumn{4}{l}{\textbf{Modification Feature (12):} $\sum|\phi|$=0.0801; $\mu|\phi|$=0.0067} \\
\hline
\rowcolor{gray!20}Author\_Prior\_Commits & 4 & 0.0235 & -0.69 \\
\rowcolor{gray!20}Commit\_Del & 9 & 0.0105 & -0.15 \\
\hline
\multicolumn{4}{l}{\textbf{Comment Feature (11):} $\sum|\phi|$=0.0959; $\mu|\phi|$=0.0010} \\
\hline
\rowcolor{gray!20}Code\_Text\_Ratio & 6 & 0.0155 & 0.89 \\
Text\_Length & 7 & 0.0133 & -0.24 \\
LDA\_Topic\_2 & 8 & 0.0116 & 0.50 \\
LDA\_Topic\_1 & 10 & 0.0102 & 0.41 \\
\hline
\end{tabular}
\end{adjustbox}
\end{revision}
\end{table}

\begin{revision}
\vspace{1ex}
\noindent
\textbf{\textit{Phase II: Model Fitting and Interpretation.}}
Since our feature set spans four dimensions, we chose a Random Forest classifier with SHAP explanations to better model and interpret the complex, non-linear interactions between features and comment addressing.
We further discuss this choice in Section~\ref{sec:threats}.
To enable focused analysis, we mapped the 3,879 comments into a binary classification task: comments labeled as ``Valid-Partially'' or ``Valid-Fully'' were grouped as the ``Addressed'' class (25.1\%), with ``Valid-Unaddressed'' comments forming the ``Not Addressed'' class (74.9\%).
Upon our refined 36 features, we trained a Random Forest classifier using stratified 5-fold cross-validation.
This approach ensures that our performance metrics and interpretability results are robust and not biased by a single random data split.
The model achieved an average of 88.6\% overall accuracy (Macro-F1=0.846) across the five test folds, indicating that our engineered features are effective predictors of comment addressing.
Since Random Forest’s native feature importance metrics lack directional insights, we applied SHAP to quantify each feature’s contribution towards ``Addressed'' predictions.
Specifically, we computed SHAP values for every instance in the test set of each fold and aggregated them to form a comprehensive explanation for the entire dataset.
SHAP is a widely-used technique to interpret learning-based models.
It explains individual predictions by assigning each feature a Shapley value, which represents its average marginal contribution to the prediction across all possible feature combinations.
Based on the aggregated SHAP values, we calculated two interpretability metrics for each feature: 1) \textbf{SHAP Importance}: the mean absolute SHAP value, reflecting global predictive importance; and 2) \textbf{SHAP Directionality}: the Pearson correlation between feature values and their SHAP values, indicating the association direction (positive or negative) with ``Addressed'' predictions.

\vspace{1ex}
\noindent
\textbf{\textit{Phase III: AI-Specific Modeling and Interpretation.}}
In addition to modeling the entire dataset, we also conducted an AI-specific modeling and interpretation to ensure that the specific factors influencing the effectiveness of AI-generated comments were not overshadowed by the human-authored ones.
To achieve this, we re-trained the Random Forest classifier using only the 2,990 AI-generated comments, removed the \textit{Is\_Human} feature, and mapped two sets of one-hot features (\textit{Trigger\_[auto/manual]} and \textit{LLM\_[GPT-3.5/4]}) into two binary features (\textit{Is\_Trigger\_Manual} and \textit{Is\_Model\_GPT4}).
The resulting AI-Specific Model achieved an overall accuracy of 91.7\% and a Macro-F1 score of 0.647 in 5-fold cross-validation.
Given the highly imbalanced nature of the dataset (only 9.2\% addressed comments), we consider this performance reasonable for supporting the subsequent interpretation analysis.
Finally, we applied the same SHAP-based analysis to the remaining 33 features to interpret the model’s predictions.
\end{revision}

\begin{table}[t]
\scriptsize
\centering
\begin{revision}
\caption{\textbf{RQ3:} \textbf{Top 10 Features} Influencing \textbf{AI Comment Addressing}. Gray-shaded rows denote features that also rank among the top 10 for total comment addressing (Table~\ref{tab:shap_feature_contribution_global}).}
\label{tab:shap_feature_contribution_ai}
\setlength{\tabcolsep}{3pt}
\begin{adjustbox}{max width=0.48\textwidth}
\renewcommand{\arraystretch}{1.1}
\begin{tabular}{c|c|c|c}
\hline
\textbf{Feature} & \textbf{Rank} & \textbf{Importance ($|\phi|$)} & \textbf{Directionality ($\rho$)} \\
\hline
\multicolumn{4}{l}{\textbf{Source Features (7):} $\sum|\phi|$=0.0418; $\mu|\phi|$=0.0060} \\
\hline
Is\_Action\_3 & 2 & 0.0177 & 0.92 \\
\rowcolor{gray!20}Is\_File\_Level\_Action & 7 & 0.0096 & -0.88 \\
\hline
\multicolumn{4}{l}{\textbf{Repository Features (3):} $\sum|\phi|$=0.0203; $\mu|\phi|$=0.0068} \\
\hline
\rowcolor{gray!20}Repo\_File\_Size & 6 & 0.0101 & 0.42 \\
\hline
\multicolumn{4}{l}{\textbf{Modification Features (12):} $\sum|\phi|$=0.0767; $\mu|\phi|$=0.0064} \\
\hline
\rowcolor{gray!20}Author\_Prior\_Commits & 1 & 0.0212 & -0.67 \\
Commit\_Base\_Lines & 4 & 0.0111 & 0.11 \\
\rowcolor{gray!20}Commit\_Del & 5 & 0.0107 & -0.17 \\
Comment\_Add & 8 & 0.0095 & 0.81 \\
\hline
\multicolumn{4}{l}{\textbf{Comment Features (11):} $\sum|\phi|$=0.0709; $\mu|\phi|$=0.0064} \\
\hline
\rowcolor{gray!20}Code\_Text\_Ratio & 3 & 0.0124 & 0.78 \\
LDA\_Topic\_5 & 9 & 0.0085 & 0.61 \\
Prior\_Comment\_Len & 10 & 0.0075 & 0.65 \\
\hline
\end{tabular}
\end{adjustbox}
\end{revision}
\end{table}

\vspace{1ex}
\noindent
\textbf{Results:}
\begin{revision}
Table~\ref{tab:shap_feature_contribution_global} presents the interpretability results based on the complete dataset (including both human and AI comments). We list the top 10 most important features, organized by their feature groups.
For each group, we also report the total SHAP Importance ($\sum|\phi|$) and average SHAP Importance ($\mu|\phi|$).
In our subsequent analysis, we mainly focus on features that show a meaningful correlation (i.e. $|\rho|\!>\!0.3$).
Notably, ``Source Features'' and ``Comment Features'' rank first and second in total importance ($\sum|\phi|$), suggesting that both \textit{who} provides the feedback (source credibility) and \textit{what} the feedback contains (content quality) are strongly associated with whether comments are addressed.
\end{revision}

\textbf{Source Features} represent the most significant group, showing a clear difference in developers' responses based on the comment origins (human vs. actions; file-level vs. hunk-level).
The negative correlation for \textit{Is\_File\_Level\_Action} ($\rho\!=\!-0.95$) aligns with the trend in Table~\ref{fig:annotation_results_detail}, where hunk-level review actions generally outperform file-level counterparts.
Specifically, \texttt{coderabbitai/ai-pr-reviewer} achieved the highest valid and addressed rate (19.2\%).
Automatic triggers also show a negative correlation with addressing (\textit{Trigger\_auto}, $\rho\!=\!-0.96$).
To detail this effect, we analyzed data distribution by trigger modes across the four actions (Table~\ref{tab:addressing_by_trigger}).
We found that in the two actions with manually triggered comments (Action ID-1 and ID-2), manually triggered comments consistently showed higher addressing rates.
This suggests that the massive feedback resulting from unconditional triggering might reduce developers' willingness to respond.
\begin{revision}
Interestingly, the LLM choice did not show high predictive importance (LLM\_GPT-3.5 ranked 19/36 and LLM\_GPT-4 ranked 21/36 in the feature list of the overall model; Is\_Model\_GPT4 ranked 23/33 in the feature list of the AI-specific model), indicating that simply employing a more advanced model does not decisively guarantee comment addressing.
However, when comparing addressing rates between two model families (Table~\ref{tab:addressing_by_llm}), GPT-4 generated comments did show better addressing rates.
\end{revision}


For \textbf{Comment Features}, we find the comments were more likely to be addressed if they were concise (\textit{Text\_Length}, $\rho\!=\!-0.24$) or contained a higher proportion of code (\textit{Code\_Text\_Ratio}, $\rho\!=\!0.89$), especially when they included multi-line code blocks(\textit{Has\_Multiline\_Code}, $\rho\!=\!0.67$ not shown in the table).
\revForMinor{As shown in Table~\ref{tab:addressing_by_code_ratio}}, the addressing rate for both human and AI-generated comments increased noticeably when \textit{Code\_Text\_Ratio} exceeded 0.5.
This may be because multi-line code is often used to illustrate concrete solutions, allowing developers to follow comments with minimal effort through copy-pasting.
Regarding topics, comments were addressed more often when they pertained to
``Concurrency Control and Locking Robustness'' (e.g., ``\textit{using a `set-if-not-exists' pattern}'') (\textit{LDA\_Topic\_2}, $\rho\!=\!0.50$),
as well as  ``UI Usability and Accessibility Enhancements'' (e.g., ``\textit{adding aria-label for screen reader support}'') (\textit{LDA\_Topic\_1}, $\rho\!=\!0.41$).

Within \textbf{Modification Features}, review comments targeting commits by experienced contributors were less likely to be addressed (\textit{Author\_Prior\_Commits}, $\rho\!=\!-0.69$), suggesting that AI-generated reviews may be more helpful for newcomers.
As shown in Table~\ref{tab:addressing_by_commit}, valid comments directed at project newcomers ($\textit{Author\_Prior\_Commits}\!\leq\!124$) achieved 16\% addressing rate, compared to just 3.3\% for the most experienced contributors.
Notably, human reviews mainly focused on these newcomers, with 79\% of comments directed at them, aligning with typical community practices.
In contrast, automated actions reviewed code changes regardless of author experience, which may help mitigate potential experience blindspots in the review process.
Finally, positive correlations for \textbf{Repository Features} demonstrated that comments were more likely to be addressed in large and active projects (\textit{Repo\_File\_Size}, $\rho\!=\!0.70$; \textit{Repo\_Issue\_Count}, $\rho\!=\!0.64$, ranked 13/36).


\begin{table}[t]
\scriptsize
\centering

\begin{revision}
\caption{\textbf{RQ3:} Addressing rates by \textbf{Trigger Mode} across review actions (\textit{p}-values from Fisher's Exact Test). }
\vspace{-2pt}
\label{tab:addressing_by_trigger}
\begin{adjustbox}{max width=0.48\textwidth}
\begin{tabular}{c|cc|cc|c}
\hline
\multirow{2}{*}{\textbf{Action}} & \multicolumn{2}{c|}{\textbf{Auto Trigger}} & \multicolumn{2}{c|}{\textbf{Manual Trigger}} & \multirow{2}{*}{\textbf{\textit{p}-value}} \\ \cline{2-5}
 & \multicolumn{1}{c|}{\textbf{Total}} & \textbf{Addr. (\%)} & \multicolumn{1}{c|}{\textbf{Total}} & \textbf{Addr. (\%)} &  \\ \hline
\textbf{Total} & \multicolumn{1}{c|}{2,886} & 9.0\% & \multicolumn{1}{c|}{104} & 14.4\% & $>$0.05 \\ \hline
\textbf{ID-1} & \multicolumn{1}{c|}{1,595} & 6.8\% & \multicolumn{1}{c|}{86} & 12.8\% & $\mathbf{\le 0.05}$ \\ \hline
\textbf{ID-2} & \multicolumn{1}{c|}{602} & 0.5\% & \multicolumn{1}{c|}{18} & 22.2\% & $\mathbf{\le 0.05}$ \\ \hline
\textbf{ID-3} & \multicolumn{1}{c|}{534} & 25.7\% & \multicolumn{1}{c|}{---} & --- & --- \\ \hline
\textbf{ID-4} & \multicolumn{1}{c|}{155} & 7.1\% & \multicolumn{1}{c|}{---} & --- & --- \\ \hline
\end{tabular}
\end{adjustbox}

\vspace{1em}

\caption{\textbf{RQ3:} Addressing rates by \textbf{LLM Series} across review actions (\textit{p}-values from Fisher's Exact Test).}
\vspace{-2pt}
\label{tab:addressing_by_llm}
\begin{adjustbox}{max width=0.48\textwidth}
\begin{tabular}{c|cc|cc|c}
\hline
\multirow{2}{*}{\textbf{Action}} & \multicolumn{2}{c|}{\textbf{GPT-3.5 Series}} & \multicolumn{2}{c|}{\textbf{GPT-4 Series}} & \multirow{2}{*}{\textbf{\textit{p}-value}} \\ \cline{2-5}
 & \multicolumn{1}{c|}{\textbf{Total}} & \textbf{Addr. (\%)} & \multicolumn{1}{c|}{\textbf{Total}} & \textbf{Addr. (\%)} &  \\ \hline
\textbf{Total} & \multicolumn{1}{c|}{1,435} & 5.4\% & \multicolumn{1}{c|}{1,555} & 12.6\% & $\mathbf{\le 0.05}$ \\ \hline
\textbf{ID-1} & \multicolumn{1}{c|}{1,431} & 5.5\% & \multicolumn{1}{c|}{250} & 16.4\% & $\mathbf{\le 0.05}$ \\ \hline
\textbf{ID-2} & \multicolumn{1}{c|}{---} & --- & \multicolumn{1}{c|}{620} & 1.1\% & --- \\ \hline
\textbf{ID-3} & \multicolumn{1}{c|}{4} & 0.0\% & \multicolumn{1}{c|}{530} & 25.8\% & $> 0.05$ \\ \hline
\textbf{ID-4} & \multicolumn{1}{c|}{---} & --- & \multicolumn{1}{c|}{155} & 7.1\% & --- \\ \hline
\end{tabular}
\end{adjustbox}

\vspace{1em}

\caption{\textbf{RQ3:} Comment \textbf{Addressing rates} by \textbf{binned Code\_Text\_Ratio} (All Four Actions vs. Human).}
\vspace{-2pt}
\label{tab:addressing_by_code_ratio}
\renewcommand{\arraystretch}{1.1}
\begin{adjustbox}{max width=0.48\textwidth}
\begin{tabular}{c|cc|cc}
\hline
\multirow{2}{*}{\makecell{\textit{Bins of}\\\textbf{Code\_Text\_Ratio}}} & \multicolumn{2}{c|}{\textbf{Four Actions}} & \multicolumn{2}{c}{\textbf{Human}} \\
\cline{2-5} 
 & \multicolumn{1}{c|}{\textbf{Total}} & \textbf{Addr. (\%)} & \multicolumn{1}{c|}{\textbf{Total}} & \textbf{Addr. (\%)} \\
\hline
(–0.00, 0.00] & \multicolumn{1}{c|}{642} & 4.2\% & \multicolumn{1}{c|}{320} & 71.2\% \\
(0.00, 0.11] & \multicolumn{1}{c|}{669} & 7.2\% & \multicolumn{1}{c|}{61} & 65.6\% \\
(0.11, 0.26] & \multicolumn{1}{c|}{635} & 5.7\% & \multicolumn{1}{c|}{95} & 75.8\% \\
(0.26, 0.52] & \multicolumn{1}{c|}{621} & 10.5\% & \multicolumn{1}{c|}{107} & 78.5\% \\
(0.52, 1.00] & \multicolumn{1}{c|}{423} & 23.2\% & \multicolumn{1}{c|}{306} & 89.9\% \\
\hline
\end{tabular}
\end{adjustbox}

\vspace{1em}

\caption{\textbf{RQ3:} Comment \textbf{Addressing rates} by \textbf{binned Author\_Prior\_Commits} (All Four Actions vs. Human).}
\vspace{-2pt}
\label{tab:addressing_by_commit}
\renewcommand{\arraystretch}{1.1}
\begin{adjustbox}{max width=0.48\textwidth}
\begin{tabular}{c|cc|cc}
\hline
\multirow{2}{*}{\makecell{\textit{Bins of}\\\textbf{Author\_Prior\_Commits}}} & \multicolumn{2}{c|}{\textbf{Four Actions}} & \multicolumn{2}{c}{\textbf{Human}} \\
\cline{2-5} 
 & \multicolumn{1}{c|}{\textbf{Total}} & \textbf{Addr. (\%)} & \multicolumn{1}{c|}{\textbf{Total}} & \textbf{Addr. (\%)} \\
\hline
(–0.00, 30] & \multicolumn{1}{c|}{218} & 16.1\% & \multicolumn{1}{c|}{558} & 80.1\% \\
(30, 124] & \multicolumn{1}{c|}{639} & 16.1\% & \multicolumn{1}{c|}{140} & 82.1\% \\
(124, 319] & \multicolumn{1}{c|}{642} & 12.1\% & \multicolumn{1}{c|}{131} & 71.8\% \\
(319, 1,013] & \multicolumn{1}{c|}{763} & 4.5\% & \multicolumn{1}{c|}{32} & 81.2\% \\
(1013, 4,316] & \multicolumn{1}{c|}{728} & 3.3\% & \multicolumn{1}{c|}{28} & 60.7\% \\
\hline
\end{tabular}
\end{adjustbox}

\vspace{-20pt}
\end{revision}

\end{table}

\begin{revision}
Table~\ref{tab:shap_feature_contribution_ai} presents the interpretability results for AI-generated comments.
As indicated by the gray-shaded rows, features that rank highly in both the total and AI-specific models demonstrate the same influence direction ($\rho$).
These consistent influence directions indicate that our main findings remain stable when restricting the analysis to AI-generated comments only.
Although trigger-mode features become less prominent here, our earlier comparison across trigger modes (Table~\ref{tab:addressing_by_trigger}) still supports higher addressing rates for manually triggered comments.
Interestingly, when focusing solely on AI-generated comments, suggestions related to ``Error Handling and Code Documentation'' (e.g., ``\textit{Instead of printing stack, use proper logging for exceptions}'')(\textit{LDA\_Topic\_5}, $\rho\!=\!0.58$) become more influential and are positively associated with addressing.
Finally, the rise of \textit{Author\_Prior\_Commits} to the top rank with a negative correlation ($\rho\!=\!-0.67$) highlights that there is still significant room for AI reviewers to evolve to become valuable for experienced developers.
\end{revision}

\begin{rqanswerbox}
\textbf{Answer to RQ3}: Comment addressing is strongly associated with \textbf{Source Features} (human vs. AI and specific action type) and \textbf{Comment Features} (conciseness and code-richness are better), highlighting the importance of thoughtful automated review design. 
\end{rqanswerbox}

\section{Discussion}
\label{sec:discussion}

\textbf{Does AI-based code review lead to code changes?} In general, human-authored review comments are much more likely to be addressed than AI-generated ones: 60\% of valid human review comments led to code changes, compared to only 0.9\%--19.2\% for valid AI-generated comments depending on the tool (Table~\ref{fig:annotation_results_detail}). However, the answer is not simply ``no''. Our results show that there is a clear potential for AI-generated code review to influence development if these tools are designed well.

\subsection{Design Matters}

Our study reveals that \textit{not all AI-based review tools are equal}. Among the four actions we analyzed in depth, one stood out: \texttt{coderabbitai/ai-pr-reviewer} achieved a 19.2\% rate of valid comments addressed, outperforming other tools by a large margin. In contrast, some tools had near zero impact. For example, \texttt{mattzcarey/code-review-gpt} had only 0.9\% of valid comments addressed. This suggests that \textit{tool design choices directly affect the developer's response}.

Through our interpretability analysis (Table~\ref{tab:shap_feature_contribution_global}), we identified several characteristics that consistently increased the likelihood of AI-generated comments being addressed.

\begin{itemize}
    \item \textbf{Hunk-level granularity:} Comments attached to specific code locations perform better than general or file-level comments. Tools such as \texttt{coderabbitai/ai-pr-reviewer} that offer hunk-level feedback are more effective.
    \item \textbf{Manual triggering:} Tools explicitly triggered by developers produce higher addressing rates. For example, \texttt{anc95/ChatGPT-CodeReview} showed a 12.8\% addressing rate for manually triggered comments versus 6.8\% for automatically triggered ones (Table~\ref{tab:addressing_by_trigger}).
    \item \textbf{Code-rich, concise comments:} Comments with a high code-to-text ratio, especially multiline suggestions, are more likely to be acted upon. We observed a clear increase in the addressing rate when more than half of the comment content was code (Table~\ref{tab:addressing_by_code_ratio}).
    \item \textbf{Focus on newcomers:} Comments were more likely to be addressed when targeting contributors with little prior commit history. For the least experienced developers, the addressing rate was five times higher than for the most experienced (Table~\ref{tab:addressing_by_commit}).
\end{itemize}

\begin{revision}
To further investigate the effectiveness of review granularity, we randomly investigated 14.4\%, i.e., 30 comments, of the 208 PR-level review comments as follows: 5 from Action ID-8, 5 from ID-11, and 20 from ID-16. Only 1 comment was partially addressed, which provided a specific performance improvement: \textit{``For better performance, preallocate the `architectures' slice by using `make([]types.Architecture, 0, len(fm.Spec.Architectures))' ''}. This aligns with our finding that concrete code suggestions are more likely to be adopted.
However, the majority of PR-level comments were ignored, supporting our conclusion that fine-grained (i.e., hunk-level) reviews are more effective for AI-generated code review.
\end{revision}

We also highlight that prompt design plays an important role in generating high-quality comments.
For example, \texttt{anc95/ChatGPT-CodeReview} uses a short, generic prompt such as ``\textit{Below is a code patch, please help me do a brief code review on it. Any bug risks and/or improvement suggestions are welcome:}'', which often results in vague comments. In contrast, \texttt{coderabbitai/ai-pr-reviewer} provides a long and detailed prompt (over 400 words), explicitly instructing the model to avoid generalities and provide focused feedback. It also includes logic to suppress comments such as ``\textit{LGTM!}'' unless no issues are found, further reducing noise.

These differences are reflected in the types of output generated. We observed many vague or unhelpful comments from tools with limited prompt context. For example, a frequent output was: ``\textit{Without more context, it is difficult to provide further suggestions}'' -- a clear signal that the model lacked sufficient information to provide value. Other typical outputs included overly generic summaries of code functionality or hallucinated style warnings unrelated to the actual changes.

\begin{figure}[t] 
    \begin{revision}
	\centering
    \setlength{\abovecaptionskip}{3pt}
	\includegraphics[width=0.49\textwidth]{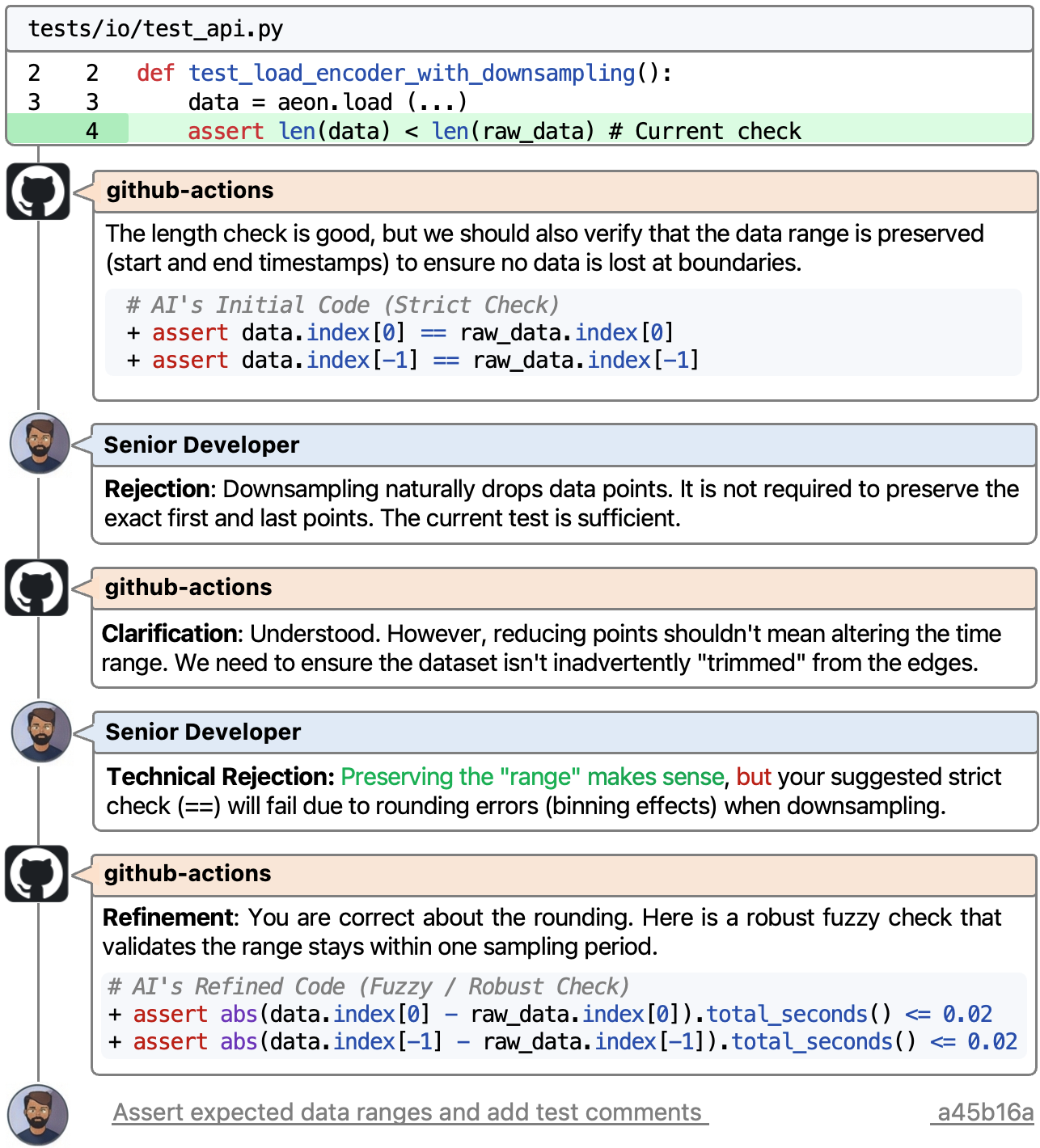} 
	\caption{Illustrative case (\href{https://github.com/SainsburyWellcomeCentre/aeon_mecha/pull/407\#discussion_r1764204980}{\textit{link}}): Effective tool design (\textbf{hunk-level review}, \textbf{concrete code suggestions}, and \textbf{interactive dialogue}) successfully persuading a senior developer to adopt a change.
    }
	\label{fig:motivation_example}
    \end{revision}
\end{figure}

To help reduce noise, some tools also implemented logic to avoid redundant reviews. For example, \texttt{coderabbitai/ai-pr-reviewer} logs reviewed commits to avoid a full base-to-head review every time.
This helps prevent developer frustration caused by redundant feedback, a concern also noted by Cihan et al.~\cite{cihan2025automated} in their study on AI-assisted code review tools in industrial settings.

\begin{revision}
We present an illustrative case in Figure~\ref{fig:motivation_example} to show why these design choices matter. This case drawn from \texttt{coderabbitai/ai-pr-reviewer}, demonstrates how concrete modifications and interactive feedback can persuade even an experienced developer (\textit{Author\_Prior\_Commits}=186) to accept a suggestion they initially rejected. After the developer identified potential issues with rounding errors, the AI adapted its proposal from a strict equality check to a robust fuzzy comparison, successfully transforming the initial rejection into an actionable improvement.
\end{revision}

Beyond post-comment refinement, we also argue that AI tools could be more cautious before posting.
We found that most current tools rely on a simplistic one-in-one-out paradigm, where each code change automatically leads to a review comment, regardless of whether a comment is warranted. This leads to low-precision output and reviewer fatigue. Instead, we advocate for a multi-stage approach, similar to practices in traditional machine learning pipelines. The tool should first determine whether a change needs review, then decide whether to generate a comment, and finally, where appropriate, offer a code suggestion. This would bring structure and discipline to AI-based review and help ensure that comments are meaningful and justified.

\begin{revision}
\subsection{Beyond Code Changes: Impact on Closed PRs and Feedback on Unaddressed Comments}
\label{sec:beyond_change}
While we mainly focused on merged PRs to observe definitive code improvements, we recognize that AI-generated review comments can provide value even when they do not lead to a merge or any direct code changes.
Therefore, we conducted two supplementary analyses on the comments outside our main dataset.

\subsubsection{Analysis of AI Comment Impact in Closed PRs}
We extended our analysis to closed (non-merged) pull requests for the four primary actions to investigate if developers engage with AI feedback before a PR is abandoned.
Following the methodology proposed in RQ2, we applied the LLM-assisted framework to assess whether comments were valid and addressed in the latest commits of these closed PRs.
We observed a significantly lower addressing rate for AI comments in closed PRs compared to merged ones: only 3.0\% (17/570) for Action ID-1, and 0\% for Action ID-2 ($n\!=\!14$), ID-3 ($n\!=\!18$), and ID-4 ($n\!=\!0$).
Our qualitative analysis suggests that even when a comment is valid, the PR may still be abandoned because the overall approach ``\texttt{doesn't solve the root cause}''.
Therefore, we focused on merged PRs in our main results to ensure we capture actual code improvements that successfully transitioned into the project’s codebase.

\subsubsection{Analysis of Feedback on Unaddressed AI Comments}
To capture impact beyond immediate code changes, we analyzed developer replies to AI comments that were not labeled as ``Valid, Addressed'' in Table~\ref{fig:annotation_results_detail}. We identified a total of 50 replies and manually categorized them into four types:
\begin{itemize}
    \item \textbf{Defending the Change (64\%):} Developers defended their current implementation as correct or preferred and explained why the AI suggestion was not adopted, e.g., ``\textit{I don't want to bring back `GetParameter' as using type parameters is a lot nicer}'' (\href{https://github.com/pokt-network/pocket/pull/622#discussion_r1160376623}{\textit{link}}).
    \item \textbf{Prompting Reflection (14\%):} AI comments triggered technical debates, which may lead to code changes depending on the outcome. For instance, a developer questioned the suggested optimization: ``\textit{Is `iterrows' really more efficient than vectorized operations? The golden rule is benchmarks}'' (\href{https://github.com/SainsburyWellcomeCentre/aeon_mecha/pull/407#discussion_r1764204980}{\textit{link}}).
    \item \textbf{Assessing Tool Capabilities (12\%):} Beyond replying to a specific suggestion, developers evaluated the overall performance of the AI or action. For instance, one developer noted it was ``\textit{interesting to learn where AI stands}'' (\href{https://github.com/pokt-network/pocket/pull/622#discussion_r1160912529}{\textit{link}}) despite the specific suggestion being of limited value.
    \item \textbf{Willingness to Follow (10\%):} Developers expressed an intention to apply the suggestion in future iterations, e.g., ``\textit{Templates could be enhanced later on to use markdown instead of raw text}'' (\href{https://github.com/gfargo/coco/pull/75#discussion_r1394481793}{\textit{link}}).
\end{itemize}
Specifically, \textbf{Willingness to Follow} (10\%) and \textbf{Prompting Reflection} (14\%) imply that even if not addressed immediately, a comment may still trigger code changes in the future.
These findings motivate future work on impact beyond code changes (e.g., guiding developer reasoning or shaping long-term code quality) to better understand the contribution of AI code reviewers to software engineering practice.
\end{revision}

\subsection{Recommended Features for Effective AI Code Review}
Based on our findings, we recommend that designers of AI-based code review tools prioritize the following.

\begin{itemize}
    \item \textbf{Use hunk-level review granularity:} Inline comments tied to specific code changes are more actionable than general or file-level comments.
    \item \textbf{Enable and encourage manual triggering:} Give developers control over when the tool runs to avoid flooding pull requests with unhelpful comments.
    \item \textbf{Include specific code suggestions:} Multi-line code blocks are especially effective as they can often be copy-pasted with minimal effort.
    \item \textbf{Filter out vague comments:} Use a pre-check (like our Stage-1 classification model) to suppress non-actionable suggestions before they reach developers.
    \item \textbf{Support adaptive reviewing:}
    \begin{itemize}
        \item Target less experienced contributors who may benefit more from feedback.
        \item Adjust the level of detail according to the change in size or file criticality.
    \end{itemize}
    \item \textbf{Provide richer context to the LLM:} Supplying the PR title, description, summaries of changes, or even user's feedback helps generate more useful review comments.
\end{itemize}

By adopting these strategies, AI-based review tools can become more than noise -- they can become useful collaborators in the development process, both in open-source communities and in industrial practice.

\subsection{Implications for AI Code Review Adopters}

We also derive the following implications for developers considering or currently adopting AI-based code review tools.

\vspace{1ex}
\noindent
\textbf{Choose and tune the tool appropriately:}
Given the large performance differences among AI-based review tools, we recommend prioritizing well-designed and highly customizable actions such as \texttt{coderabbitai/ai-pr-reviewer} as a starting point.
Second, do not rely solely on the default settings. In our study, 82.6\% of the repositories analyzed customized their action configurations, and 22.5\% continued to adjust the configurations beyond one month. This suggests that integrating AI code review tools requires careful tuning and ongoing optimization. Based on our observations in RQ1, this includes practices such as keeping the action and its underlying LLMs up to date, and addressing project-specific needs by customizing prompts, refining file inclusion/exclusion patterns, and adjusting trigger modes.

\vspace{1ex}
\noindent
\textbf{Clarify the role of AI code review:}
Even the best-performing AI tool achieved only a 19.2\% addressing rate, far below the human reviewers' 60\%.
Therefore, for now, AI code review is better framed as a \textit{complement} to human review, not a replacement.
This requires understanding AI code review's capability boundaries and deploying it in appropriate scenarios. 
Based on our LDA topic analysis (RQ3), AI reviewers are good at providing defensive programming suggestions, and developers are relatively more responsive to their comments on error handling. 
Accordingly, we recommend positioning AI reviewers as \textit{defenders}: tightly scope them to this mandate (with targeted prompts and other recommended features), thereby suppressing category-sprawling yet low-quality comments.
This allows human reviewers to focus on higher-level issues, such as overall architecture (e.g. ``\textit{Try sorting at the DB level, it will be 100x faster}''), development context (e.g., ``\textit{Remove this as project requires Python~$\geq$~3.11}''), and business logic (e.g., ``\textit{This will cause issues on ARM64}'').
Mapping the boundaries of AI code review's effectiveness is also an ongoing task for researchers.\looseness=-1

\vspace{1ex}
\noindent
\textbf{Monitor effectiveness over time:} We recommend using the LLM-assisted framework proposed in this study to monitor comment validity and addressing rates over time. These metrics provide a crucial feedback loop to continually tune and customize the AI reviewer.

\subsection{Implications for GitHub}

For GitHub, we recommend the following to better support the development and analysis of automated code review tools.

\vspace{1ex}
\noindent
\textbf{Enhance the traceability of Actions' behaviors:}
Currently, GitHub retains action execution logs for up to 400 days, and all actions operate under a shared \texttt{github-action[bot]} account. This makes it difficult to attribute specific comments or behaviors to individual actions—a threat also noted in Section~\ref{sec:threats}. Therefore, we suggest refining action identification to the particular action within a repository, which enables precise source attribution for users and researchers. Additionally, we suggest that GitHub establish a ``usage list'' (maybe labeled by a successful run) for Marketplace actions, similar to repository forks. This would also enable action developers, potential users, and researchers to assess their true popularity.

\vspace{1ex}
\noindent
\textbf{Improve support for identifying actual code changes:}
Currently, most tools re-review the full base-to-head diff every time, since there’s no easy way to trace the reviewed commits to calculate incremental changes. Therefore, we suggest GitHub provide a platform-level solution for this, rather than leaving it to individual action developers. This may be implemented through extending the existing ``viewerViewedState'' flag (a boolean indicating file view status) to the specific commit hash of the last reviewed version. Additionally, a true ``compare file content'' endpoint (that ignores noise from rebases) could also ease implementing code review tools.

\section{Threats to Validity}
\label{sec:threats}

\noindent
\textbf{Construct Validity:}
We identify AI generated comments through the \texttt{github-action[bot]} account, which may include comments from unrelated actions.
Although our manual checks during annotation revealed no such cases, misclassification remains possible.
Additionally, we determined comment addressing based on file-level code changes, which may be insufficient for certain edge cases.
However, ``uncertain'' cases constituted only 1.3\% of our addressing dataset, and we excluded these from our impact factor analysis.

\vspace{1ex}
\noindent
\textbf{Internal Validity:}
In RQ2, we extracted 150 comments from over 22,000 AI-generated and 1,166 human-authored comments and obtained gold-standard labels via multi-rater annotation. To assess representativeness, the first author examined an additional 250 samples and observed similar addressing patterns as reported in our main findings. Therefore, we believe the gold standard reflects the broader dataset.
For RQ3, our addressing analysis focuses on four popular actions, which may limit the generalizability of our findings.
However, within the scope of our dataset, usage was highly concentrated on these four actions, making them representative of real-world deployments observed in our study.
Moreover, the final analysis uses a binary classification of whether comments were addressed, which does not capture the degree or impact of the resulting code changes.
Nevertheless, our evaluation shows substantial inter-rater agreement and strong LLM performance on this task, ensuring the reliability of our findings.
Our feature engineering for impact factor analysis may also be incomplete or miss latent feature interactions.
However, our features was comprehensively designed across four dimensions and inspired by prior works~\cite{rahman2017predicting,DBLP:conf/esem/DeyM20}.
Another threat is that different interpretation methods may yield different analysis results. 
Since our feature set spans four dimensions, we chose Random Forest to better model the complex and non-linear interactions between features, rather than the simpler model like logistic regression, which assumes a linear relationship between features.
We also tried logistic regression in our experiments and it achieved a lower performance (86.1\% accuracy and 0.826 Macro-F1 in 5-fold cross-validation) than our Random Forest model (88.6\% and 0.846 respectively). 
This model choice also aligns with work~\cite{rahman2017predicting,DBLP:conf/esem/DeyM20}. 
And the use of SHAP is a standard approach for interpreting such models as well. Therefore, our interpretation pipeline is well-justified.
We also acknowledge that these interpretations describe associations between features and the predicted probability of ''Addressed'', not causal effects. 
If causal identification is desired, further studies (e.g., controlled experiments or developer interviews) would be needed.

\vspace{1ex}
\noindent
\textbf{External Validity:}
\begin{revision}
Our study includes 16 AI-based code review actions identified through a systematic Marketplace search and manual verification. Rather than sampling a subset, we included all actions that met our relevance criteria and collected the complete set of observable review comments generated by these tools in mature repositories. Although four tools account for the majority of comments (most notably \texttt{anc95/ChatGPT-CodeReview}), this concentration reflects actual ecosystem usage patterns at the time of data collection.
We acknowledge that less popular or newly released tools are underrepresented in absolute volume. However, their limited usage inherently results in fewer observable review behaviors. Thus, our dataset captures the near-complete population of AI-based review activity occurring on GitHub Actions at the time of the study, particularly among tools with meaningful adoption.
To further verify dataset coverage, we conducted a follow-up Marketplace search on July 23, 2025. We observed no substantial changes among the top-ranked actions. The only notable newcomer, \href{https://github.com/hustcer/deepseek-review}{\texttt{hustcer/deepseek-review}} (ranked \#7, 324 stars), had its first commit on January 29, 2025, i.e., after our action selection date (January 7, 2025). This confirms that our dataset accurately reflects the state of the ecosystem during the study period.
Consequently, our findings, such as the benefits of hunk-level granularity and concrete code suggestions, are derived from analysis across multiple widely used tools and human reviews, rather than from a single-tool artifact. We will continue to monitor the evolution of the AI-based code review ecosystem in future work.
\end{revision}
Second, to mitigate the impact of inactive repositories which may not yet have established stable review workflows, we followed prior research~\cite{DBLP:conf/icse/ReichM23} and collected data only from repositories with at least 50 PRs.
However, this threshold does not fully capture the diversity of project scales. 
Since our dataset primarily consists of small- to medium-sized projects (typically with $\le$ 50 non-bot contributors), our findings may not generalize to very large-scale industrial systems.
\begin{revisionForMinor}
Moreover, our main analysis focuses on review comments generated by file-level and hunk-level review actions and examines whether they are addressed in merged PRs.
We did not consider PR-level actions.
This design choice simplifies the analysis by ensuring that we can reliably observe whether comments have been addressed, but it may also narrow the generalizability of our findings.
To broaden our understanding beyond this core scope, we additionally examined a sample of PR-level review comments and conducted the supplementary analyses on closed PRs and feedback on unaddressed AI comments, presented in Section~\ref{sec:beyond_change}.
\end{revisionForMinor}
Additionally, our analysis focused on English-language comments.
As described in RQ2 Phase I, we began with 16{,}762 comments from 100 repositories.
A language filter reduced this to 4{,}229 comments from 53 repositories, excluding about 75\% of comments and 47 repositories.
Among them, 32 repositories were excluded because they only contained Korean comments (e.g., 1{,}828 comments from \href{https://github.com/KEA-DoKebi/dalkom-backend/pull/101#discussion_r1457056514}{KEA-DoKebi/dalkom-backend}).
Despite this reduction, we retained over half of the original projects.
And given that English serves as the main communication language in global development activities~\cite{7321207}, with even non-native speakers commonly adopting it, we believe our findings remain relevant to the broader open-source community.



\section{Related Work}

We organize related work into two areas relevant to our study: (a) the automation of code review using AI, and (b) developer response to automated feedback.

\paragraph{AI for Code Review Automation}

The automation of code review has attracted growing attention with the rise of GenAI. Modern large language models, such as GPT-4 and LLaMA, have shown strong performance on code-related tasks, including vulnerability detection~\cite{linevul}, program repair~\cite{vulrepair,outofsight,vitrepair}, and bug fixing~\cite{tufanobugfix,cure,jin2023inferfix}, enabled by the naturalness of source code~\cite{ray2016naturalness,hindle2016naturalness,allamanis2018survey}.

Review automation builds on the observation that many review comments are repetitive and can be learned. Prior work has proposed systems for three core tasks~\cite{lin2025codereviewqa}: (1) determining whether a code change needs review~\cite{codereviewer,hellendoorn2021towards}, (2) generating natural language comments~\cite{codereviewer,tufano2022using,li2022auger,Lin2024}, and (3) generating revised code based on comments~\cite{lin2023towards,tufano2022using,thongtanunam2022autotransform}.

Comment generation is particularly complex, requiring the synthesis of context-sensitive suggestions. Early approaches used encoder-decoder models such as T5~\cite{tufano2022using}, and subsequent work explored specialized pretraining~\cite{li2022auger,codereviewer}, cross-task learning~\cite{sghaier2024improving}, and benchmark datasets~\cite{Lin2024,lin2023cct5}. Other work has demonstrated that general-purpose models like LLaMA can be adapted to this task through fine-tuning~\cite{lu2023llama}.

\begin{revisionForMinor}
Recent work explored how LLMs can be more effectively applied to code review, e.g. by incorporating pull request context~\cite{zhang2025laura} or through mixture-of-prompts architectures~\cite{peng2025icodereviewer}. 
Meanwhile, benchmark-oriented studies also developed more realistic evaluation settings for AI-based code review~\cite{zhang2026code,pereira2026cr}. 
Additionally, Cihan et al. studied the impact of AI-assisted code review in industrial settings~\cite{cihan2025automated}.

Although some studies include automatic and human evaluations of generated comments~\cite{tufano2022using}, most of this research has focused on comment quality rather than downstream impact. 
Our study complements these efforts by analyzing how AI-generated comments are used in open-source development workflows, whether they lead to code changes, and which factors influence their effectiveness.
\end{revisionForMinor}

\paragraph{Developer Response to Automated Feedback}
Several studies have explored how developers respond to code review comments, with a focus on human-authored feedback. Prior work has investigated factors that make comments useful~\cite{bosu2015characteristics,rahman2017predicting,turzo2024makes}, developer responsiveness~\cite{macleod2017code}, and differences in comment granularity and presentation~\cite{bacchelli2013expectations,lin2024leveraging}.

Building on this foundation, more recent work has turned to developers’ reactions to AI-generated artifacts. For example, Vaithilingam et al.~\cite{vaithilingam2022expectation} examined developers’ expectations and experiences with GitHub Copilot, and Huang et al.~\cite{DBLP:journals/tosem/HuangGDSCLZZ23} studied perceptions of AI-generated inline versus block comments.

However, little is known about whether developers actually act on AI-generated code review comments. Our study addresses this gap by analyzing over 22,000 such comments across 178 repositories, assessing whether they were addressed, and identifying the factors that influence their effectiveness.

\section{Conclusion and Future Work}
This paper presents a large-scale empirical analysis of AI-based code review actions on GitHub.
We examine how these tools are adopted, whether their comments lead to code changes, and what factors influence their effectiveness.
Our findings reveal that adoption is highly concentrated on four popular actions.
Many AI-generated comments are not addressed, especially when they are vague or lack context.
However, comments that are concise, specific are more likely to be addressed—particularly by less experienced contributors.
\begin{revision}
Moreover, even when AI comments do not lead to immediate code changes, we find that they can still prompt developer reflection and inform future improvements.
\end{revision}
Based on our findings, we offer implications for code review tool designers, adopters, and GitHub.

We recommend treating AI code reviewers as collaborators with human reviewers.
Future work should further enhance review quality by exploring richer context, better prompts, and multi-stage strategies to generate more specific and actionable feedback. 
It is also essential to clarify AI's capabilities to identify optimal ways for human-AI collaboration.
We hope our findings contribute to the development of more practical and developer-friendly AI-based code review tools.

\balance
\clearpage

\section*{Acknowledgments}
This work is supported by the Fundamental and Interdisciplinary Disciplines Breakthrough Plan of the Ministry of Education of China (No.~JYB2025XDXM118), the National Natural Science Foundation of China (No.~62572237, No.~62302210, No.~72371125), the 2024 Development and Testing Tools Project (CEIEC-2024-ZM02-0066), the Frontier Technologies R\&D Program of Jiangsu (No.~BF2024059), and the Natural Science Foundation of Jiangsu Province (No.~BK20241195). 

\bibliographystyle{IEEEtran}
\bibliography{refs}

@inproceedings{DBLP:conf/icse/ReichM23,
  author       = {Pavel Reich and
                  Walid Maalej},
  title        = {Testability Refactoring in Pull Requests: Patterns and Trends},
  booktitle    = {45th {IEEE/ACM} International Conference on Software Engineering, {ICSE} 2023, Melbourne, Australia, May 14-20, 2023},
  pages        = {1508--1519},
  publisher    = {{IEEE}},
  year         = {2023},
}

@inproceedings{DBLP:conf/msr/Treude019,
  author       = {Christoph Treude and
                  Markus Wagner},
  title        = {Predicting good configurations for GitHub and stack overflow topic
                  models},
  booktitle    = {Proceedings of the 16th International Conference on Mining Software
                  Repositories, {MSR} 2019, 26-27 May 2019, Montreal, Canada},
  pages        = {84--95},
  publisher    = {{IEEE} / {ACM}},
  year         = {2019},
}

@inproceedings{DBLP:conf/icsm/JiarpakdeeTT18,
  author       = {Jirayus Jiarpakdee and
                  Chakkrit Tantithamthavorn and
                  Christoph Treude},
  title        = {AutoSpearman: Automatically Mitigating Correlated Software Metrics
                  for Interpreting Defect Models},
  booktitle    = {2018 {IEEE} International Conference on Software Maintenance and Evolution,
                  {ICSME} 2018, Madrid, Spain, September 23-29, 2018},
  pages        = {92--103},
  publisher    = {{IEEE} Computer Society},
  year         = {2018},
}

@inproceedings{DBLP:conf/nips/LundbergL17,
  author       = {Scott M. Lundberg and
                  Su{-}In Lee},
  title        = {A Unified Approach to Interpreting Model Predictions},
  booktitle    = {Advances in Neural Information Processing Systems 30: Annual Conference
                  on Neural Information Processing Systems 2017, December 4-9, 2017,
                  Long Beach, CA, {USA}},
  pages        = {4765--4774},
  year         = {2017},
}

@article{blei2003latent,
  title={Latent dirichlet allocation},
  author={Blei, David M and Ng, Andrew Y and Jordan, Michael I},
  journal={Journal of machine Learning research},
  volume={3},
  number={Jan},
  pages={993--1022},
  year={2003}
}

@inproceedings{bacchelli2013expectations,
  title={Expectations, outcomes, and challenges of modern code review},
  author={Bacchelli, Alberto and Bird, Christian},
  booktitle={2013 35th International Conference on Software Engineering (ICSE)},
  pages={712--721},
  year={2013},
  organization={IEEE}
}

@inproceedings{sadowski2018modern,
  title={Modern code review: a case study at google},
  author={Sadowski, Caitlin and S{\"o}derberg, Emma and Church, Luke and Sipko, Michal and Bacchelli, Alberto},
  booktitle={Proceedings of the 40th international conference on software engineering: Software engineering in practice},
  pages={181--190},
  year={2018}
}

@inproceedings{gousios2016work,
  title={Work practices and challenges in pull-based development: The contributor's perspective},
  author={Gousios, Georgios and Storey, Margaret-Anne and Bacchelli, Alberto},
  booktitle={Proceedings of the 38th international conference on software engineering},
  pages={285--296},
  year={2016}
}

@inproceedings{lu2023llama,
  title={Llama-reviewer: Advancing code review automation with large language models through parameter-efficient fine-tuning},
  author={Lu, Junyi and Yu, Lei and Li, Xiaojia and Yang, Li and Zuo, Chun},
  booktitle={2023 IEEE 34th International Symposium on Software Reliability Engineering (ISSRE)},
  pages={647--658},
  year={2023},
  organization={IEEE}
}

@article{wessel2023github,
  title={Github actions: the impact on the pull request process},
  author={Wessel, Mairieli and Vargovich, Joseph and Gerosa, Marco A and Treude, Christoph},
  journal={Empirical Software Engineering},
  volume={28},
  number={6},
  pages={131},
  year={2023},
  publisher={Springer}
}

@inproceedings{decan2022use,
  title={On the use of github actions in software development repositories},
  author={Decan, Alexandre and Mens, Tom and Mazrae, Pooya Rostami and Golzadeh, Mehdi},
  booktitle={2022 IEEE International Conference on Software Maintenance and Evolution (ICSME)},
  pages={235--245},
  year={2022},
  organization={IEEE}
}

@inproceedings{bosu2015characteristics,
  title={Characteristics of useful code reviews: An empirical study at microsoft},
  author={Bosu, Amiangshu and Greiler, Michaela and Bird, Christian},
  booktitle={2015 IEEE/ACM 12th Working Conference on Mining Software Repositories},
  pages={146--156},
  year={2015},
  organization={IEEE}
}

@article{turzo2024makes,
  title={What makes a code review useful to opendev developers? an empirical investigation},
  author={Turzo, Asif Kamal and Bosu, Amiangshu},
  journal={Empirical Software Engineering},
  volume={29},
  number={1},
  pages={6},
  year={2024},
  publisher={Springer}
}

@inproceedings{rahman2017predicting,
  title={Predicting usefulness of code review comments using textual features and developer experience},
  author={Rahman, Mohammad Masudur and Roy, Chanchal K and Kula, Raula G},
  booktitle={2017 IEEE/ACM 14th International Conference on Mining Software Repositories (MSR)},
  pages={215--226},
  year={2017},
  organization={IEEE}
}

@article{lin2024leveraging,
  title={Leveraging Reviewer Experience in Code Review Comment Generation},
  author={Lin, Hong Yi and Thongtanunam, Patanamon and Treude, Christoph and Godfrey, Michael W and Liu, Chunhua and Charoenwet, Wachiraphan},
  journal={arXiv preprint arXiv:2409.10959},
  year={2024}
}

@article{macleod2017code,
  title={Code reviewing in the trenches: Challenges and best practices},
  author={MacLeod, Laura and Greiler, Michaela and Storey, Margaret-Anne and Bird, Christian and Czerwonka, Jacek},
  journal={IEEE Software},
  volume={35},
  number={4},
  pages={34--42},
  year={2017},
  publisher={IEEE}
}

@inproceedings{linevul,
  title={Linevul: A transformer-based line-level vulnerability prediction},
  author={Fu, Michael and Tantithamthavorn, Chakkrit},
  booktitle={Proceedings of the 19th International Conference on Mining Software Repositories},
  pages={608--620},
  year={2022}
}

@inproceedings{vulrepair,
  title={VulRepair: a T5-based automated software vulnerability repair},
  author={Fu, Michael and Tantithamthavorn, Chakkrit and Le, Trung and Nguyen, Van and Phung, Dinh},
  booktitle={Proceedings of the 30th ACM joint european software engineering conference and symposium on the foundations of software engineering},
  pages={935--947},
  year={2022}
}

@inproceedings{outofsight,
  title={Out of sight, out of mind: Better automatic vulnerability repair by broadening input ranges and sources},
  author={Zhou, Xin and Kim, Kisub and Xu, Bowen and Han, DongGyun and Lo, David},
  booktitle={Proceedings of the IEEE/ACM 46th International Conference on Software Engineering},
  pages={1--13},
  year={2024}
}

@article{vitrepair,
  title={Vision transformer inspired automated vulnerability repair},
  author={Fu, Michael and Nguyen, Van and Tantithamthavorn, Chakkrit and Phung, Dinh and Le, Trung},
  journal={ACM Transactions on Software Engineering and Methodology},
  volume={33},
  number={3},
  pages={1--29},
  year={2024},
  publisher={ACM New York, NY, USA}
}

@article{tufanobugfix,
  title={An empirical study on learning bug-fixing patches in the wild via neural machine translation},
  author={Tufano, Michele and Watson, Cody and Bavota, Gabriele and Penta, Massimiliano Di and White, Martin and Poshyvanyk, Denys},
  journal={ACM Transactions on Software Engineering and Methodology (TOSEM)},
  volume={28},
  number={4},
  pages={1--29},
  year={2019},
  publisher={ACM New York, NY, USA}
}

@inproceedings{cure,
  title={Cure: Code-aware neural machine translation for automatic program repair},
  author={Jiang, Nan and Lutellier, Thibaud and Tan, Lin},
  booktitle={2021 IEEE/ACM 43rd International Conference on Software Engineering (ICSE)},
  pages={1161--1173},
  year={2021},
  organization={IEEE}
}

@inproceedings{jin2023inferfix,
  title={Inferfix: End-to-end program repair with llms},
  author={Jin, Matthew and Shahriar, Syed and Tufano, Michele and Shi, Xin and Lu, Shuai and Sundaresan, Neel and Svyatkovskiy, Alexey},
  booktitle={Proceedings of the 31st ACM Joint European Software Engineering Conference and Symposium on the Foundations of Software Engineering},
  pages={1646--1656},
  year={2023}
}

@inproceedings{ray2016naturalness,
  title={On the" naturalness" of buggy code},
  author={Ray, Baishakhi and Hellendoorn, Vincent and Godhane, Saheel and Tu, Zhaopeng and Bacchelli, Alberto and Devanbu, Premkumar},
  booktitle={Proceedings of the 38th International Conference on Software Engineering},
  pages={428--439},
  year={2016}
}

@article{hindle2016naturalness,
  title={On the naturalness of software},
  author={Hindle, Abram and Barr, Earl T and Gabel, Mark and Su, Zhendong and Devanbu, Premkumar},
  journal={Communications of the ACM},
  volume={59},
  number={5},
  pages={122--131},
  year={2016},
  publisher={ACM New York, NY, USA}
}

@article{allamanis2018survey,
  title={A survey of machine learning for big code and naturalness},
  author={Allamanis, Miltiadis and Barr, Earl T and Devanbu, Premkumar and Sutton, Charles},
  journal={ACM Computing Surveys (CSUR)},
  volume={51},
  number={4},
  pages={1--37},
  year={2018},
  publisher={ACM New York, NY, USA}
}

@inproceedings{codereviewer,
  title={Automating code review activities by large-scale pre-training},
  author={Li, Zhiyu and Lu, Shuai and Guo, Daya and Duan, Nan and Jannu, Shailesh and Jenks, Grant and Majumder, Deep and Green, Jared and Svyatkovskiy, Alexey and Fu, Shengyu and others},
  booktitle={Proceedings of the 30th ACM Joint European Software Engineering Conference and Symposium on the Foundations of Software Engineering},
  pages={1035--1047},
  year={2022}
}

@inproceedings{hellendoorn2021towards,
  title={Towards automating code review at scale},
  author={Hellendoorn, Vincent J and Tsay, Jason and Mukherjee, Manisha and Hirzel, Martin},
  booktitle={Proceedings of the 29th ACM Joint Meeting on European Software Engineering Conference and Symposium on the Foundations of Software Engineering},
  pages={1479--1482},
  year={2021}
}

@inproceedings{tufano2022using,
  title={Using pre-trained models to boost code review automation},
  author={Tufano, Rosalia and Masiero, Simone and Mastropaolo, Antonio and Pascarella, Luca and Poshyvanyk, Denys and Bavota, Gabriele},
  booktitle={Proceedings of the 44th international conference on software engineering},
  pages={2291--2302},
  year={2022}
}

@inproceedings{li2022auger,
  title={AUGER: automatically generating review comments with pre-training models},
  author={Li, Lingwei and Yang, Li and Jiang, Huaxi and Yan, Jun and Luo, Tiejian and Hua, Zihan and Liang, Geng and Zuo, Chun},
  booktitle={Proceedings of the 30th ACM Joint European Software Engineering Conference and Symposium on the Foundations of Software Engineering},
  pages={1009--1021},
  year={2022}
}

@inproceedings{Lin2024,
  title={Improving automated code reviews: Learning from experience},
  author={Lin, Hong Yi and Thongtanunam, Patanamon and Treude, Christoph and Charoenwet, Wachiraphan},
  booktitle={Proceedings of the 21st International Conference on Mining Software Repositories},
  pages={278--283},
  year={2024}
}

@inproceedings{lin2023towards,
  title={Towards automated code reviews: Does learning code structure help?},
  author={Lin, Hong Yi and Thongtanunam, Patanamon},
  booktitle={2023 IEEE International Conference on Software Analysis, Evolution and Reengineering (SANER)},
  pages={703--707},
  year={2023},
  organization={IEEE}
}

@inproceedings{thongtanunam2022autotransform,
  title={Autotransform: Automated code transformation to support modern code review process},
  author={Thongtanunam, Patanamon and Pornprasit, Chanathip and Tantithamthavorn, Chakkrit},
  booktitle={Proceedings of the 44th international conference on software engineering},
  pages={237--248},
  year={2022}
}

@article{sghaier2024improving,
  title={Improving the learning of code review successive tasks with cross-task knowledge distillation},
  author={Ben Sghaier, Oussama and Sahraoui, Houari},
  journal={Proceedings of the ACM on Software Engineering},
  volume={1},
  number={FSE},
  pages={1086--1106},
  year={2024},
  publisher={ACM New York, NY, USA}
}

@inproceedings{lin2023cct5,
  title={Cct5: A code-change-oriented pre-trained model},
  author={Lin, Bo and Wang, Shangwen and Liu, Zhongxin and Liu, Yepang and Xia, Xin and Mao, Xiaoguang},
  booktitle={Proceedings of the 31st ACM Joint European Software Engineering Conference and Symposium on the Foundations of Software Engineering},
  pages={1509--1521},
  year={2023}
}

@inproceedings{vaithilingam2022expectation,
  title={Expectation vs. experience: Evaluating the usability of code generation tools powered by large language models},
  author={Vaithilingam, Priyan and Zhang, Tianyi and Glassman, Elena L},
  booktitle={Chi conference on human factors in computing systems extended abstracts},
  pages={1--7},
  year={2022}
}

@inproceedings{cihan2025automated,
  title={Automated code review in practice},
  author={Cihan, Umut and Haratian, Vahid and {\.I}{\c{c}}{\"o}z, Arda and G{\"u}l, Mert Kaan and Devran, {\"O}mercan and Bayendur, Emircan Furkan and U{\c{c}}ar, Baykal Mehmet and T{\"u}z{\"u}n, Eray},
  booktitle={2025 IEEE/ACM 47th International Conference on Software Engineering: Software Engineering in Practice (ICSE-SEIP)},
  pages={425--436},
  year={2025},
  organization={IEEE}
}

@misc{Claude, title = {Claude}, year={2026}, howpublished = {\url{https://claude.ai}} }

@misc{OpenAI, title = {OpenAI}, year={2026}, howpublished = {\url{https://chat.openai.com}} }

@misc{DeepSeek, title = {DeepSeek}, year={2026}, howpublished = {\url{https://deepseek.com}} }

@misc{Gemini, title = {Gemini}, year={2026}, howpublished = {\url{https://gemini.google.com/}} }

@article{DBLP:journals/tosem/HuangGDSCLZZ23,
  author       = {Yuan Huang and
                  Hanyang Guo and
                  Xi Ding and
                  Junhuai Shu and
                  Xiangping Chen and
                  Xiapu Luo and
                  Zibin Zheng and
                  Xiaocong Zhou},
  title        = {A Comparative Study on Method Comment and Inline Comment},
  journal      = {{ACM} Trans. Softw. Eng. Methodol.},
  volume       = {32},
  number       = {5},
  pages        = {126:1--126:26},
  year         = {2023},
  url          = {https://doi.org/10.1145/3582570},
  doi          = {10.1145/3582570},
  bibsource    = {dblp computer science bibliography, https://dblp.org}
}

@inproceedings{DBLP:conf/esem/DeyM20,
  author       = {Tapajit Dey and
                  Audris Mockus},
  title        = {Effect of Technical and Social Factors on Pull Request Quality for
                  the {NPM} Ecosystem},
  booktitle    = {{ESEM} '20: {ACM} / {IEEE} International Symposium on Empirical Software
                  Engineering and Measurement, Bari, Italy, October 5-7, 2020},
  pages        = {11:1--11:11},
  publisher    = {{ACM}},
  year         = {2020},
  url          = {https://doi.org/10.1145/3382494.3410685},
  doi          = {10.1145/3382494.3410685},
  bibsource    = {dblp computer science bibliography, https://dblp.org}
}

@inproceedings{golzadeh2020bot,
  title={Bot or not? Detecting bots in GitHub pull request activity based on comment similarity},
  author={Golzadeh, Mehdi and Legay, Damien and Decan, Alexandre and Mens, Tom},
  booktitle={Proceedings of the IEEE/ACM 42nd international conference on software engineering workshops},
  pages={31--35},
  year={2020}
}

@misc{appendix,
	author = {},
	title = {Does {AI} Code Review Lead to Code Changes? {A} Case Study of {GitHub} Actions ({Online Appendix})},
	howpublished = {\url{https://zenodo.org/records/19562450}},
	year = {2026},
	doi = {10.5281/zenodo.19562449},
	note = {Zenodo, accessed 14-04-2026},
}

@misc{hunk,
	author = {},
	title = {Definition of ``Hunk'' in the GNU Diffutils Manual},
	howpublished = {\url{https://www.gnu.org/software/diffutils/manual/html_node/Hunks.html}},
	year = {2025},
}

@inproceedings{llmguidelines-paper,
    author      = {Stefan Wagner and
                  Marvin Mu{\~{n}}oz Bar{\'{o}}n and
                  Davide Falessi and
                  Sebastian Baltes},
    title       = {Towards Evaluation Guidelines for Empirical Studies involving LLMs},
    booktitle   = {2nd International Workshop on Methodological Issues with Empirical Studies in Software Engineering (WSESE 2025)},
    year = 2025
}

@inproceedings{lin2025codereviewqa,
  title={The Code Review Comprehension Assessment for Large Language Models},
  author={Lin, Hong Yi and Liu, Chunhua and Gao, Haoyu and Thongtanunam, Patanamon and Treude, Christoph},
   booktitle={Findings of the Association for Computational Linguistics: ACL 2025},
  year={2025}
}

@INPROCEEDINGS{7321207,
  author={Wang, Yi},
  booktitle={2015 ACM/IEEE International Symposium on Empirical Software Engineering and Measurement (ESEM)}, 
  title={Language Matters}, 
  year={2015},
  volume={},
  number={},
  pages={1-10},
  doi={10.1109/ESEM.2015.7321207}}

@misc{baltes2025evaluationguidelinesempiricalstudies,
      title={Evaluation Guidelines for Empirical Studies in Software Engineering involving LLMs}, 
      author={Sebastian Baltes and Florian Angermeir and Chetan Arora and Marvin Muñoz Barón and Chunyang Chen and Lukas Böhme and Fabio Calefato and Neil Ernst and Davide Falessi and Brian Fitzgerald and Davide Fucci and Marcos Kalinowski and Stefano Lambiase and Daniel Russo and Mircea Lungu and Lutz Prechelt and Paul Ralph and Christoph Treude and Stefan Wagner},
      year={2025},
      eprint={2508.15503},
      archivePrefix={arXiv},
      primaryClass={cs.SE},
      url={https://arxiv.org/abs/2508.15503}, 
}

@inproceedings{zhang2025laura,
  title={Laura: Enhancing code review generation with context-enriched retrieval-augmented llm},
  author={Zhang, Yuxin and Zhang, Yuxia and Sun, Zeyu and Jiang, Yanjie and Liu, Hui},
  booktitle={2025 40th IEEE/ACM International Conference on Automated Software Engineering (ASE)},
  pages={2983--2995},
  year={2025},
  organization={IEEE}
}

@inproceedings{peng2025icodereviewer,
  title={icodereviewer: Improving secure code review with mixture of prompts},
  author={Peng, Yun and Kim, Kisub and Meng, Linghan and Liu, Kui},
  booktitle={2025 40th IEEE/ACM International Conference on Automated Software Engineering (ASE)},
  pages={3204--3215},
  year={2025},
  organization={IEEE}
}

@article{zhang2026code,
  title={Code Review Agent Benchmark},
  author={Zhang, Yuntong and Pan, Zhiyuan and Yusuf, Imam Nur Bani and Ruan, Haifeng and Shariffdeen, Ridwan and Roychoudhury, Abhik},
  journal={arXiv preprint arXiv:2603.23448},
  year={2026},
  url={https://arxiv.org/pdf/2603.23448}
}

@article{pereira2026cr,
  title={CR-Bench: Evaluating the Real-World Utility of AI Code Review Agents},
  author={Pereira, Kristen and Sinha, Neelabh and Ghosh, Rajat and Dutta, Debojyoti},
  journal={arXiv preprint arXiv:2603.11078},
  year={2026},
  url={https://arxiv.org/pdf/2603.11078}
}


 




\vfill

\end{document}